\documentclass[pre,aps,twocolumn,superscriptaddress]{revtex4-1}
\usepackage{graphicx}
\usepackage{amssymb,amsfonts,amsmath}
\usepackage{color}
\usepackage{ulem}
\usepackage[hidelinks]{hyperref}
\usepackage{bbm}
\usepackage{bbold}

\newcommand{\rv}{{\mathbf r}}

\newcommand{\Tr}{{\rm Tr}\,}
\newcommand{\e}{{\rm e}}

\newcommand{\Jv}{{\bf J}}

\newcommand{\pv}{{\bf p}}

\newcommand{\Fv}{{\bf F}}
\newcommand{\fv}{{\bf f}}

\newcommand{\vel}{{\bf v}}

\newcommand{\ms}{\color{black}}

\newcommand{\msphantom}[1]{$\ldots$}

\newcommand{\eps}{{\boldsymbol \epsilon}}

\newcommand{\rt}{(\rv,t)}

\newcommand{\eqr}[1]{Eq.~\eqref{#1}}

\newcommand{\avg}[1]{\Big\langle {\protect #1} \Big\rangle}

\newcommand{\mydelete}[1]{{}}

\newcommand{\rmint}{{\rm int}}
\newcommand{\rmad}{{\rm ad}}
\newcommand{\rmsup}{{\rm sup}}
\newcommand{\rmexc}{{\rm exc}}
\newcommand{\rmext}{{\rm ext}}

\newcommand{\rmid}{{\rm id}}
\newcommand{\x}{\rv}
\renewcommand{\star}{{}}

\begin{document}

\title{Why neural functionals suit statistical mechanics}

\author{Florian Samm\"uller}
\affiliation{Theoretische Physik II, Physikalisches Institut, 
  Universit{\"a}t Bayreuth, D-95447 Bayreuth, Germany}

\author{Sophie Hermann}
\affiliation{Theoretische Physik II, Physikalisches Institut, 
  Universit{\"a}t Bayreuth, D-95447 Bayreuth, Germany}

\author{Matthias Schmidt}
\affiliation{Theoretische Physik II, Physikalisches Institut, 
  Universit{\"a}t Bayreuth, D-95447 Bayreuth, Germany}
\email{Matthias.Schmidt@uni-bayreuth.de}

\date{26 August 2023}
\date{29 November 2023, {\ms revised version: 14 February 2024}}

\begin{abstract}
  We describe recent progress in the statistical mechanical
  description of many-body systems via machine learning combined with
  concepts from density functional theory and many-body
  simulations. We argue that the neural functional theory by
  Samm\"uller {\it et
    al.}\ [\href{https://doi.org/10.1073/pnas.2312484120}{Proc.\ Natl.\ Acad. Sci. {\bf
        120}, e2312484120 (2023)}] gives a functional representation
  of direct correlations and of thermodynamics that allows for
  thorough quality control and consistency checking of the involved
  methods of artificial intelligence. Addressing a prototypical system
  we here present a pedagogical application to hard core particle in
  one spatial dimension, where Percus' exact solution for the free
  energy functional provides an unambiguous reference. A corresponding
  standalone numerical tutorial that demonstrates the neural
  functional concepts together with the underlying fundamentals of
  Monte Carlo simulations, classical density functional theory,
  machine learning, and differential programming is available online
  at \href{https://github.com/sfalmo/NeuralDFT-Tutorial}
  {https://github.com/sfalmo/NeuralDFT-Tutorial}.
\end{abstract}

\maketitle

\section{Introduction}
\label{SECintroduction}

The discovery of the molecular structure of matter was still in its
infancy when van der Waals predicted in 1893 on theoretical grounds
that the gas-liquid interface has finite thickness. The theory is
based on a square-gradient treatment of the density inhomogeneity
between the coexisting phases \cite{vanderwaals1893,
  RowlinsonWidomBook} and it is consistent with van der Waals' earlier
treatment of the gas-liquid phase separation in bulk. Both the bulk
and the interfacial treatments are viewed as simple yet physically
correct descriptions of fundamental phase coexistence phenomena by
modern standards of statistical mechanics.

What was unknown then is that an underlying formally exact variational
principle exists. This mathematical structure was recognized only much
later, first quantum mechanically by Hohenberg and Kohn
\cite{hohenberg1964} for the groundstate of a many-body system,
subsequently by Mermin \cite{mermin1965} for finite temperatures, and
then classically by Evans~\cite{evans1979}. The variational principle
forms the core of density functional theory and the intervening
history between the quantum~\cite{mermin1965} and classical milestones
\cite{evans1979} is described by Evans {\it et al.}~\cite{evans2016};
much background of the theory is given in Refs.~\cite{evans1992,
  hansen2013, schmidt2022rmp}.  Kohn and Sham \cite{kohn1965,
  kohn1999} re-introduced orbitals via an effective single-particle
description, which facilitates the efficient treatment of the
many-electron quantum problem.

Practical applications of density functional theory require one to
make concrete approximations for the central functional.  (We recall
that a functional maps an entire function to a number.)  Quantum
mechanically one needs to approximate the exchange-correlation energy
functional $E_{\rm xc}[n]$, as depending on the electronic density
profile $n(\rv)$, and classically one needs to get to grips with the
excess (over ideal gas) intrinsic Helmholtz free energy
$F_\rmexc[\rho]$, as a functional of the local particle
density~$\rho(\rv)$.

A broad range of relevant problems and intriguing collective and
self-organization effects in soft matter \cite{evans2019physicsToday}
have been investigated on the basis of classical density functional
theory \cite{evans1979, evans1992, evans2016, hansen2013,
  schmidt2022rmp}. Exemplary topical studies include investigations of
hydrophobicity \cite{levesque2012jcp, evans2019pnas, coe2022prl,
  jeanmairet2013jcp}, the orientation-resolved molecular structure of
liquids \cite{jeanmairet2013jcp}, the three-dimensionally resolved
atomic structure of electrolytes \cite{martinjimenez2017natCom,
  hernandez-munoz2019}, and the asymptotic decay of ionic structural
correlations \cite{cats2021decay}.

Owing to its rigorous formal foundation, density functional theory
provides a microscopic, first-principles treatment of the many-body
problem. The numerical efficiency of (in practice often approximate)
implementations allows for exhaustive model parameter sweeps, for
systematic investigation of bulk and interfacial phase transitions,
and for the discovery and tracing of scaling laws. Exact statistical
mechanical sum rules \cite{baus1984, evans1990, henderson1992,
  upton1998} integrate themselves very naturally into the scheme and
they provide consistency checks and can form the basis for refined
approximations. Nevertheless, at the core of such studies lies usually
an approximate functional and hence resorting to explicit many-body
simulations is common in a quest for validation of the predicted
density functional results.

Inline with topical developments in other branches of science, the use
of machine learning is becoming increasingly popular in soft matter
research. Recent applications of machine learning range from the
characterization of soft matter \cite{clegg2021ml},
reverse-engineering of colloidal self-assembly \cite{dijkstra2021ml},
local structure detection in colloidal systems \cite{boattinia2019ml},
to the investigation of many-body potentials for isotropic
\cite{campos2021ml} and for anisotropic \cite{campos2022ml} colloids.
Brief overviews of machine learning in physics \cite{rodrigues2023}
and in particular in liquid state theory \cite{wu2023review} were
given recently.

Density functional theory lends itself towards machine learning due
the necessity of finding an approximation for the central functional.
Corresponding research was carried out in the classical
\cite{teixera2014, lin2019ml, lin2020ml, cats2022ml, qiao2020,
  yatsyshin2022, malpica-morales2023, fang2022, simon2023mlPatchy,
  delasheras2023perspective, sammueller2023neural,
  sammueller2023neuralTutorial} and quantum realms \cite{nagai2018,
  jschmidt2018, zhou2019, nagai2020, li2021prl, li2022natcompsci,
  gedeon2022, pederson2022, huang2023review}. The classical work
addressed liquid crystals in complex confinement \cite{teixera2014},
the functional construction of a convolutional network
\cite{lin2019ml} and of an equation-learning network \cite{lin2020ml},
the improvement of the standard mean-field approximation for the
three-dimensional Lennard-Jones system \cite{cats2022ml} with the aim
of addressing gas solubility in nanopores \cite{qiao2020}, the use
physics-informed Bayesian inference \cite{yatsyshin2022,
  malpica-morales2023}, active learning with error
control~\cite{fang2022}, and the physics of patchy particles
\cite{simon2023mlPatchy}.

The quantum mechanical problem was addressed on the basis of machine
learning the exchange-correlation potential \cite{nagai2018,
  jschmidt2018, zhou2019}, testing its out-of-training transferability
\cite{nagai2018}, using a three-dimensional convolutional neural
network construct \cite{zhou2019}, considering hidden messages from
molecules \cite{nagai2020}, and using the Kohn-Sham equations already
during training via a regularizer method \cite{li2021prl}.  The
Hamiltonian itself was targeted via deep learning with the aim of
efficient electronic-structure calculation \cite{li2022natcompsci}. A
recent perspective on these and more developments was given by Burke
and co-workers \cite{pederson2022}. Huang {\it et
  al.}\ \cite{huang2023review} argue prominently that quantum density
functional theory plays a special role in the wider context of the use
of artificial intelligenece methods in chemistry and in materials
science.

While the central problem of quantum density functional theory is to
deal with the exchange and correlation effects between electrons that
are exposed to the external field generated by the nuclei, classical
statistical mechanics of soft matter relies on a much more varied
range of underlying model Hamiltonians. The effective interparticle
interactions in soft matter systems cover a wide gamut of different
types of repulsive and attractive, short- and long-ranged, hard-,
soft-, and penetrable-core behaviours.

\begin{figure}[!t]
  \includegraphics[page=1,width=.99\columnwidth]{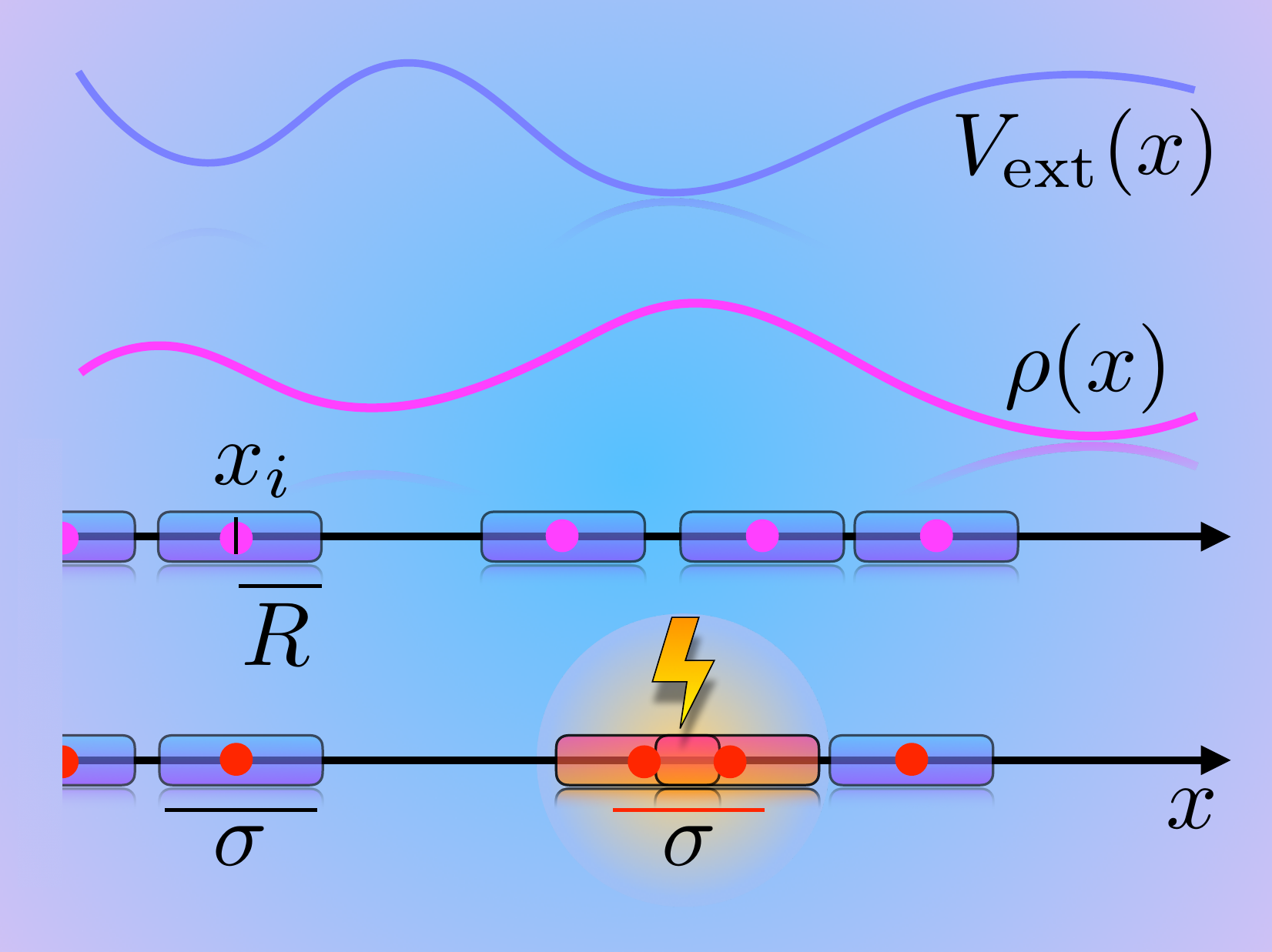}
\caption{Illustration of hard rods in one spatial dimension that are
  exposed to a position-dependent external potential $V_\rmext(x)$. In
  response to the external influence a spatially inhomogeneous density
  profile $\rho(x)$ emerges in equilibrium at temperature $T$ and
  chemical potential $\mu$. The particles with position
  coordinates~$x_i$ and particle index $i=1,\ldots, N$ have radius $R$
  and diameter $\sigma=2R$. A configuration is forbidden (bottom row)
  if any two particles overlap, i.e., if their mutual distance is
  smaller than the particle diameter $\sigma$.}
\label{FIGrods}
\end{figure}

In particular the hard core model plays a special role. For hard core
particles the pair potential between two particles is infinite if the
particle pair overlaps and it vanishes otherwise. Hard core particles
are relatively simple as temperature becomes an irrelevant variable
while the essence of short-ranged repulsion and the resulting
molecular packing remain captured correctly \cite{santos2020review,
  royall2023review}. The statistical mechanics of the bulk of
one-dimensional hard core particles was solved early by Tonks
\cite{tonks1936}. The free energy functional is known exactly due to
Percus \cite{percus1976, robledo1981, vanderlick1989, bakhti2012,
  percus2013} and his solution provides the general structure and
thermodynamics of the system when exposed to an external potential,
see Fig.~\ref{FIGrods} for an illustration.  The mathematical form of
Percus' free energy functional was one of the sources of inspiration
\cite{rosenfeld1988} for Rosenfeld's powerful fundamental measure
density functional for three-dimensional hard spheres
\cite{rosenfeld1989, tarazona2000, roth2002WhiteBear,
  roth2006WhiteBear, roth2010, kierlik1990, kierlik1991,
  phan1993}. One-dimensional hard rods are also central for
nonequilibrium physics~\cite{marconi1999, lips2018, lips2019,
  lips2020, antonov2022} and the Percus functional forms a highly
useful reference for developing and testing machine learning
techniques in classical density functional theory \cite{lin2019ml,
  lin2020ml, fang2022, yatsyshin2022, malpica-morales2023}.

In recent work, de las Heras {\it et
  al.}\ \cite{delasheras2023perspective} and Samm\"uller {\it et
  al.}\ \cite{sammueller2023neural} have put forward machine learning
strategies that operate on the one-body level of correlation
functions.  Here we address in detail the neural functional theory
\cite{sammueller2023neural} for inhomogeneous fluids in equilibrium.
We argue that this approach constitutes a neural network-based theory,
where multiple different and mutually intimately related neural
functionals form a genuine theoretical structure that permits
investigation, testing, and to ultimately gain profound insight into
the nature of the coupled many-body physics.  Thereby the training is
only required for a single neural network, from which then all further
neural functionals are created in straightforward ways.  The method
allows for multi-scale application \cite{sammueller2023neural} as is
pertinent for many areas of soft matter~\cite{schmid2022editorial,
  baptista2021, gholami2021}.  It is furthermore applicable to general
interactions, as exemplified by successfully addressing a
supercritical Lennard-Jones fluid \cite{sammueller2023neural}, thus
complementing analytical efforts to construct density functional
approximations. Such work was based, e.g., on hierarchical integral
equations \cite{yagi2021, yagi2022}, on functional renormalization
group methods \cite{iso2019densityRenormalization, yokota2021,
  kawana2023}, and on fundamental measure theory
\cite{schmidt1999sfmt,schmidt2000sfmtMix,schmidt2000sfmtStructure}.

Here we use the one-dimensional hard core model to illustrate the key
concepts of the neural functional theory, as the required sampling can
be performed easily and Percus' functional provides an analytical
structure that we can relate to the neural theory.  The Percus
functional is one of the very few general classical free energy
density functionals that is analytically known for a continuum model
(see e.g.\ also Refs.~\cite{percus1982, buschle2000}) and this fact
provides further motivation for our study.  A hands-on tutorial that
demonstrates the key concepts of constructing a neural direct
correlation functional, generating the required data from Monte Carlo
simulations, testing against a numerical implementation of the Percus
functional, and working with automatic differentiation is available
online~\cite{sammueller2023neuralTutorial}.

The paper is structured into individual subsections, as described in
the following; each subsection is self-contained to a significant
degree such that Readers are welcome to select the description of
those topics that match their own interests and individual
backgrounds.  An overview of key concepts of the one-body neural
functional approach is given in Sec.~\ref{SECoverview}. This hybrid
method draws on classical density functional concepts, as summarized
in Sec.~\ref{SECdft}. Functional differentiation and integration
methods are described in Sec.~\ref{SECfunctionalCalculus}.

Readers who are primarily interested in the use of machine learning
may want to skip the above material and rather start with
Sec.~\ref{SECtraining}, where we describe how to construct and train
the neural correlation functional on the basis of many-body simulation
data. We concentrate on the specific model of one-dimensional hard
core particles and complement and contrast the neural functional by
the known exact analytical results for this model, as described in
Sec.~\ref{SECpercusFunctional}.  Model applications for predicting
inhomogeneous systems based on neural density functional theory are
described in Sec.~\ref{SECapplication}.

Several methods of neural functional calculus are described in
Sec.~\ref{SECneuralCalculus}. Manipulating the neural correlation
functional by functional integration and automatic functional
differentiation is described in Secs.~\ref{SECintegrating} and
\ref{SECdifferentiation}, respectively.  The application of Noether
sum rules as a standalone means for quality control of the neural
network is presented in Sec.~\ref{SECnoether}. Functional integral sum
rules are shown in Sec.~\ref{SECfunctionalIntegralSumRules}.  A brief
overview of key concepts of neural functional representations in
nonequilibrium are presented in Sec.~\ref{SECnonequilibrium}.  We give
conclusions in Sec.~\ref{SECconclusions}.

\subsection{Neural functional concepts}
\label{SECoverview}

\begin{figure}[!t]
  \includegraphics[page=1,width=.99\columnwidth]{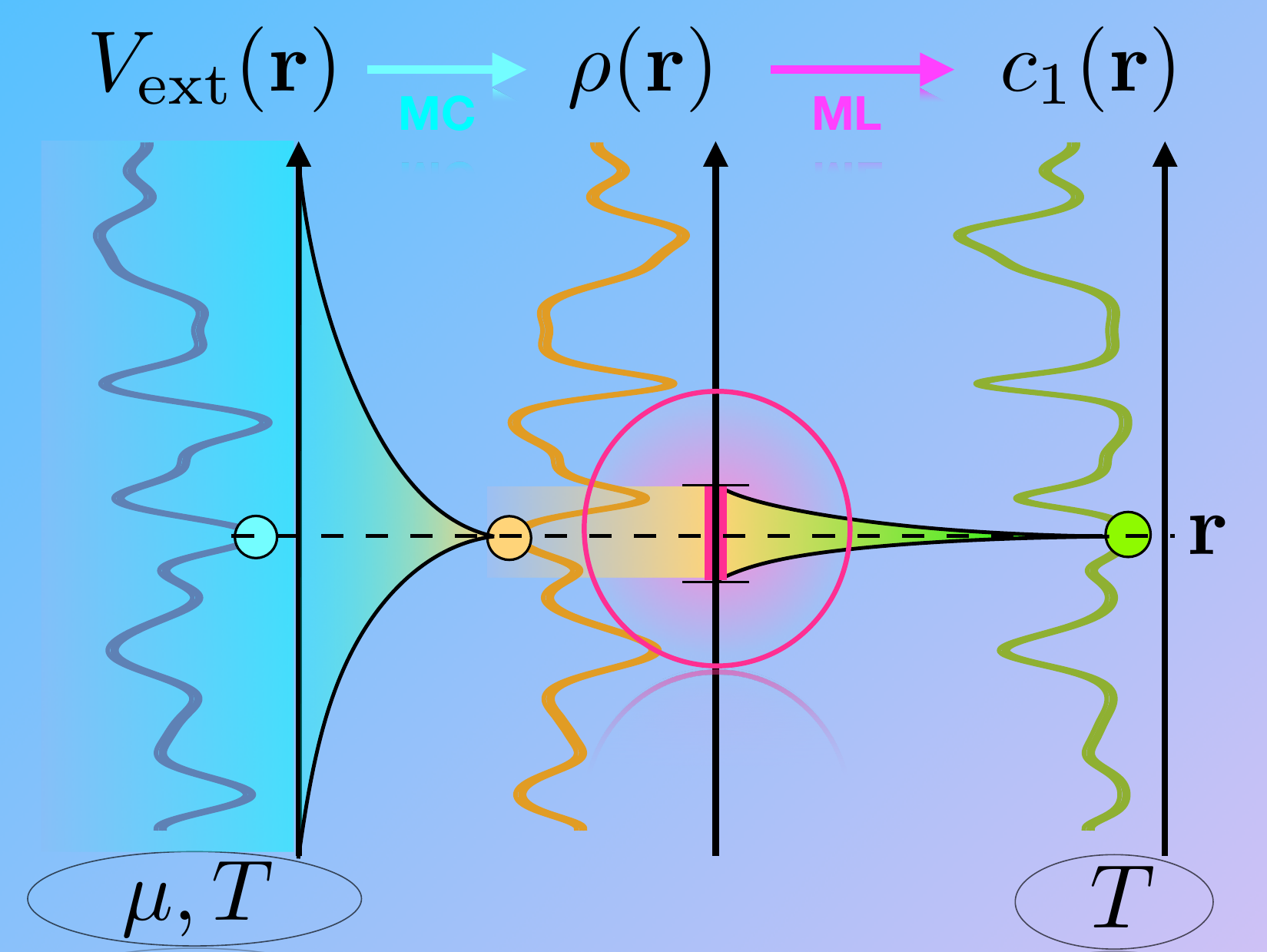}
\caption{Illustration of the relevant functional maps of the neural
  functional theory.  The external potential $V_\rmext(\rv)$ generates
  a one-body density profile $\rho(\rv)$ that is associated with a
  one-body direct correlation function $c_1(\rv)$. At given
  temperature $T$, chemical potential $\mu$, and for a specific form
  of the external potential $V_\rmext(\rv)$, Monte Carlo simulations
  provide data for the corresponding density profile $\rho(\rv)$ and
  for the direct correlation function $c_1(\rv)$. Machine learning is
  used to represent the functional map $\rho\to c_1$ via a deep neural
  network. The functional dependence of $c_1(\rv)$ on the density
  profile is of much shorter spatial range as compared to the training
  data obtained from $V_\rmext\to \rho$.}
\label{FIGnftOverview}
\end{figure}

The neural functional framework \cite{sammueller2023neural} rests on a
combination of simulation, density functional theory, and machine
learning. Data that characterizes the underlying many-body system is
generated via grand canonical Monte Carlo simulations of well-defined,
but random external conditions. Based on these results the one-body
direct correlation functional is constructed as a neural network that
accepts as an input the relevant local section of the density
profile. This method allows for very efficient data handling as only
short-ranged correlations contribute; Fig.~\ref{FIGnftOverview}
depicts an illustration.

The neural one-body direct correlation functional $c_1(\rv,[\rho])$
forms the mother network for the subsequent functional
calculus. Automatically differentiating the mother network with
respect to its density input yields the two-body direct correlation
functional $c_2(\rv,\rv',[\rho])$ as a daughter functional. Two-body
direct correlations are central in liquid state theory
\cite{hansen2013} and they are here represented by a standalone
numerical object that is created via straightforward application of
automatic differentiation. This workflow is very different and
arguabaly much simpler in practice than the standard technique of
carrying out the functional differentiation analytically and then
implementing the resulting expression(s) via numerical code.

Differentiating the daughter network yields a granddaughter network,
which represents the three-body direct correlation functional
$c_3(\rv,\rv',\rv'',[\rho])$.  Again this is an independent and
standalone numerical computing object. Very little is known about
three-body direct correlations, with e.g.\ Rosenfeld's early
investigation for hard spheres \cite{rosenfeld1989} and the freezing
studies by Likos and Ashcroft \cite{likos1992, likos1993} being
notable exceptions.  The neural functional
method~\cite{sammueller2023neural} offers arguably unprecedented
detailed access.

Tracing the genealogy in the reverse direction requires functional
integration, which is a general and standard technique in functional
calculus. In the present case again a quasi-standalone numerical
object can be built based on mere network evaluation and standard
numerical integration, both of which are fast operations. In this way
functionally integrating the mother one-body direct correlation
functional creates as the grandmother the excess free energy
functional $F_\rmexc[\rho]$. This mathematical object is the ultimate
generating functional in classical density functional theory for all
$n$-body {\it direct} correlation functions \cite{evans1979,
  hansen2013, schmidt2022rmp}. We give more details about the
interrelationships within the family of functionals below in
Sec.~\ref{SECfunctionalCalculus}.

When applied to the three-dimensional hard sphere fluid and restricted
to planar geometry, such that the density distribution is
inhomogeneous only along a single spatial direction, the neural
functional theory outperforms the best available hard sphere density
functional (the formidable White Bear Mk.\ II fundamental measure
theory \cite{roth2010}) in generic inhomogeneous situations. For
spatially homogeneous fluids the neural functional even surpasses the
``very accurate equation of state'' \cite{hansen2013} by Carnahan and
Starling \cite{santos2020review}, despite the fact that no explicit
information about {\it any} bulk fluid properties was used during
training.

Formulating reliable strategies of how to test machine-learning
predictions constitutes in general a complex yet very important task,
not least in the light of ongoing and projected increased use of
artificial intelligence in science \cite{huang2023review}. The neural
functional theory offers a wealth of concrete self-consistency checks
besides the standard benchmarking techniques. Commonly and following
best practice in machine learning, benchmarking is performed by
dividing the reference data, as here obtained from many-body
simulations, into training, validation and test data. The simulations
in the test data set have not been used during training and hence can
serve to assess the performance of the trained network.  In our
present model application, we can perform testing directly with
respect to the exact Percus theory.

Assessing extrapolation capabilities beyond the underlying data set
requires the availability of further reference data.  In
Ref.~\cite{sammueller2023neural} this is provided by comparing
(favourably) against a highly accurate bulk equation of state
\cite{kolafa2004} as well as comparing against free energy reference
results obtained from simulation-based thermodynamic integration of
inhomogeneous systems.

However, due to its computational efficiency the neural approach
allows to make predictions for system sizes that outscale
significantly the dimensions of the original simulation
box. Ref.~\cite{sammueller2023neural} describes systems of
micron-sized colloids confined between parallel walls with macroscopic
separation distance. The density profile is resolved over a system
size of 1\,mm with nanometric precision on a numerical grid with
10\,nm spacing. Such ``simulation beyond the box'' is both powerful in
terms of multiscale description of soft matter
\cite{schmid2022editorial, baptista2021, gholami2021}, but is also
serves as template for the more general situation of using artificial
intelligence methods far outside their original training realm.

In order to provide quality control, the neural functional theory
hence allows to carry out a second type of test. This is less generic
than the above benchmarking but it can nevertheless provide
inspiration for machine learning in wider contexts. In the present
case, the specific statistical mechanical nature of the underlying
equilibrium many-body system implies far-reaching mathematical
structure, as it lies at the very heart of Statistical
Mechanics. Specifically, it is the significant body of equilibrium sum
rules that provide formally exact interrelations between different
types of correlation functions. These sum rules hold universally,
i.e., independent of the specific inhomogeneous situation that is
under consideration and they hence constitute formally exact
relationships between functionals.

As the neural functional theory expresses direct correlation functions
using neural network methods, the sum rules directly translate to
identities that connect the different neural functionals and their
integrated and differentiated relatives with each other. Crucially,
these connections have both different mathematical form, as well as
different physical meaning, as compared to the bare genealogy provided
by the automatic functional differentiation and functional
integration. Without overstretching the analogy, one could view the
sum rules as genetic testing the entire family for absence of
inheritable disease.

While the body of statistical mechanical sum rules is both significant
and diverse \cite{baus1984, evans1990, henderson1992, upton1998}, here
we rely on the recent Noether invariance theory
\cite{hermann2021noether, hermann2022topicalReview,
  hermann2022variance, tschopp2022forceDFT, sammueller2022forceDFT,
  hermann2022quantum, sammueller2023whatIsLiquid, robitschko2023any,
  hermann2023whatIsLiquid} as a systematic means to create both known
and new functional identities from the thermal invariance of the
underlying statistical mechanics \cite{hermann2021noether,
  hermann2022topicalReview}. In particular from invariance against
local shifting one obtains sum rules that connect different
generations of direct correlation functionals with each other in both
locally-resolved and global form. We present exemplary cases below in
Sec.~\ref{SECnoether}. Generic sum rules that emerge from the mere
inverse relationship of functional integration and functional
differentation are presented in
Sec.~\ref{SECfunctionalIntegralSumRules}.

\subsection{Introduction to classical density functional theory}
\label{SECdft}
We give a compact account of some key concepts of classical density
functional theory; for more details see Refs.~\cite{evans1979,
  evans1992, evans2016, hansen2013, schmidt2022rmp}. Readers who are
primarily interested in machine learning of neural functionals can
skip this and the next subsection and directly proceed to
Sec.~\ref{SECnft}.

In a statistical mechanical description of a many-body system the
local density acts as a generic order parameter that measures the
probability of finding a particle at a specific location. The formal
definition of the one-body density distribution as a statistical
average is:
\begin{align}
  \rho(\rv) &= \avg{\sum_i\delta(\rv-\rv_i)},
  \label{EQrhoAverage}
\end{align}
where the sum over $i$ runs over all $N$ particles, $\rv_i$ is the
position coordinate of particle $i=1,\ldots, N$, and $\delta(\cdot)$
indicates the Dirac distribution, here in three dimensions. The angles
indicate a thermal average over microstates, which can e.g.\ be
efficiently carried out in Monte Carlo simulations.

For completeness, we give a formal description of the equilibrium
average based on the grand ensemble, where it is defined as $\langle
\cdot \rangle = \Tr \cdot \e^{-\beta (H-\mu N)}/\Xi$. Here the inverse
temperature is $\beta=1/(k_BT)$, with the Boltzmann constant $k_B$ and
absolute temperature $T$, the Hamiltonian $H$, chemical potential
$\mu$ and grand partition sum $\Xi$. The classical trace is defined as
$\Tr \cdot = \sum_{N=0}^\infty (h^{3N}N!)^{-1} \int d\rv^N \int d\pv^N
\cdot$, where $h$ denotes the Planck constant and $\int d\rv^N \int
d\pv^N$ is a shorthand for the high-dimensional phase space integral
over all particle positions and momenta. Pedagogical introductions can
be found in standard textbooks \cite{hansen2013} and an introductory
compact account together with a description of the force point of view
is provided in Ref.~\cite{hermann2022topicalReview}.

The Hamiltonian has the following standard form:
\begin{align}
  H = \sum_i \frac{\pv_i^2}{2m} + u(\rv^N) + \sum_i V_\rmext(\rv_i),
  \label{EQhamiltonian}
\end{align}
where $\pv_i$ is the momentum of particle $i$, the interparticle
interaction potential $u(\rv^N)$ depends on all position coordinates
$\rv^N=\rv_1,\ldots,\rv_N$, and $V_\rmext(\rv)$ is an external
potential energy function that depends on position $\rv$. Hence the
sum in \eqr{EQhamiltonian} comprises kinetic, interparticle, and
external energy contributions. For the common case of particles
interacting via a pair potential $\phi(r)$ that only depends on the
interparticle distance $r$, the interparticle energy reduces to
$u(\rv^N)=\sum_{ij(\neq)} \phi(|\rv_i-\rv_j|)/2$ where the double sum
runs only over distinct particle pairs $ij$ with $i\neq j$ and the
factor $1/2$ corrects for double counting.

For the ideal gas the interparticle interactions vanish,
$u(\rv^N)\equiv 0$, and the density profile is given by the
generalized barometric law \cite{hansen2013}:
\begin{align}
  \rho_\rmid(\rv)=\e^{-\beta (V_\rmext(\rv)-\mu)}/ \Lambda^d,
  \label{EQrhoideal}
\end{align}
where $\Lambda$ denotes the thermal de Broglie wavelength, which in
the present classical case can be set $\Lambda=\sigma$, with $\sigma$
denoting the particle size; for simplicity of notation here we use
$\Lambda=1$; we have indicated the spatial dimensionality by $d$.

Taking the logarithm of \eqr{EQrhoideal} and collecting all terms on
the left hand side gives the following ideal gas chemical potential
balance:
\begin{align}
  \ln\rho_\rmid(\rv) + \beta V_\rmext(\rv)- \beta \mu &= 0.
  \label{EQrhoidealLog}
\end{align}

For a mutually interacting system, where $u(\rv^N)\neq 0$,
\eqr{EQrhoidealLog} will not be true when replacing the ideal density
profile $\rho_\rmid(\rv)$ by the true density profile $\rho(\rv)$ as
formally given by \eqr{EQrhoAverage}. Rather the sum of the three
terms on the left hand side of \eqr{EQrhoidealLog} will not vanish,
but yield a nontrivial contribution:
\begin{align}
  \ln \rho(\rv) + \beta V_\rmext(\rv) - \beta \mu &= c_1(\rv),
  \label{EQel}
\end{align}
where the one-body direct correlation function $c_1(\rv)$ is in
general nonzero and arises due to the presence of interparticle
interactions in the system. (For hard core systems $c_1(\rv)$
typically features negative values.)

The machine learning strategy described below in
Sec.~\ref{SECtraining} is based on this pragmatic access to data for
$c_1(\rv)$, as obtained by direct simulation of $\rho(\rv)$ on the
basis of explicitly carrying out the average in \eqr{EQrhoAverage} for
given form of $V_\rmext(\rv)$ and prescribed values of the
thermodynamic parameters $\mu$ and~$T$. As the one-body direct
correlation function is central in the neural functional theory, we
combine Eqs.~\eqref{EQrhoidealLog} and \eqref{EQel}, which yields the
following equivalent form for the one-body direct correlation
function,
\begin{align}
  c_1(\rv) = \ln\Big(\frac{\rho(\rv)}{\rho_\rmid(\rv)}\Big),
  \label{EQc1AsRatio}
\end{align}
where $\rho_\rmid(\rv)$ is given by \eqr{EQrhoidealLog} with
$\Lambda=1$. Equation~\eqref{EQc1AsRatio} has the direct
interpretation of $c_1(\rv)$ as the logarithm of the ratio of the
actual density profile and the density profile of the ideal gas under
identical conditions, as given by the external potential and
thermodynamic statepoint.

In alternative terminology~\cite{hansen2013} one defines the intrinsic
chemical potential as $\mu_\rmint(\rv)=\mu - V_\rmext(\rv)$. The
intrinsic chemical potential and the one-body direct correlation
function are related trivially to each other via $\mu_\rmint(\rv)=
k_BT [\ln\rho(\rv) - c_1(\rv)]$ as is obtained straightforwardly by
re-arranging \eqr{EQel}.

The practical, computational, and conceptual advantage of density
functional theory lies in avoiding the explicit occurrence of the
high-dimensional phase space integral that underlies thermal averages;
we recall the definition of the density profile \eqref{EQrhoAverage}
as such an expectation value. Instead, and without any principal loss
of information, one works with functional dependencies. Rather than
mere point-wise dependencies, such as between the functions
$\rho(\rv)$, $V_\rmext(\rv)$, and $c_1(\rv)$ that hold at each
point~$\rv$, see \eqr{EQel}, a functional dependence is on the
entirety of a function and it has in general a nonlocal and nonlinear
structure.

Density functional theory is specifically based on the fact
\cite{hohenberg1964, mermin1965, evans1979} that for a given type of
fluid, as characterized by its interparticle interaction potential
$u(\rv^N)$, and known thermodynamic parameters $\mu$ and $T$, the form
of density profile $\rho(\rv)$ is sufficient to determine the entirety
of the external potential $V_\rmext(\rv)$. Hence a unique functional
map exists \cite{hohenberg1964, mermin1965, evans1979}:
\begin{align}
  \rho \to V_\rmext.
  \label{EQfunctionalMap}
\end{align}
Here we omit the position arguments on both sides to reflect in the
notation that the functional map relates the entirety of the density
profile to the entirety of the external potential.

Applying \eqr{EQfunctionalMap} to the external potential, as it occurs
in \eqr{EQel}, implies that the left hand side is determined from
knowledge of the density profile alone, in principle without any need
for a priori knowledge of the form of $V_\rmext(\rv)$. Via the
identity \eqr{EQel} we can conclude the existence of the map:
\begin{align}
  \rho \to c_1,
\end{align}
where the entirety of the density profile determines the entirety of
the direct correlation function.  As a consequence the one-body direct
correlation function actually is a density functional,
$c_1(\rv,[\rho])$, where the brackets indicate the functional
dependence, i.e.\ on the entirety of the argument function, here
$\rho(\rv)$. We will discuss below more explicitly that the dependence
is effectively short-ranged for the case of short-ranged interparticle
interaction potentials and that this can be exploited to great effect
in the neural network methodology.

\subsection{Density functional derivatives and integrals}
\label{SECfunctionalCalculus}

\begin{figure}[!t]
  \includegraphics[page=1,width=.99\columnwidth]{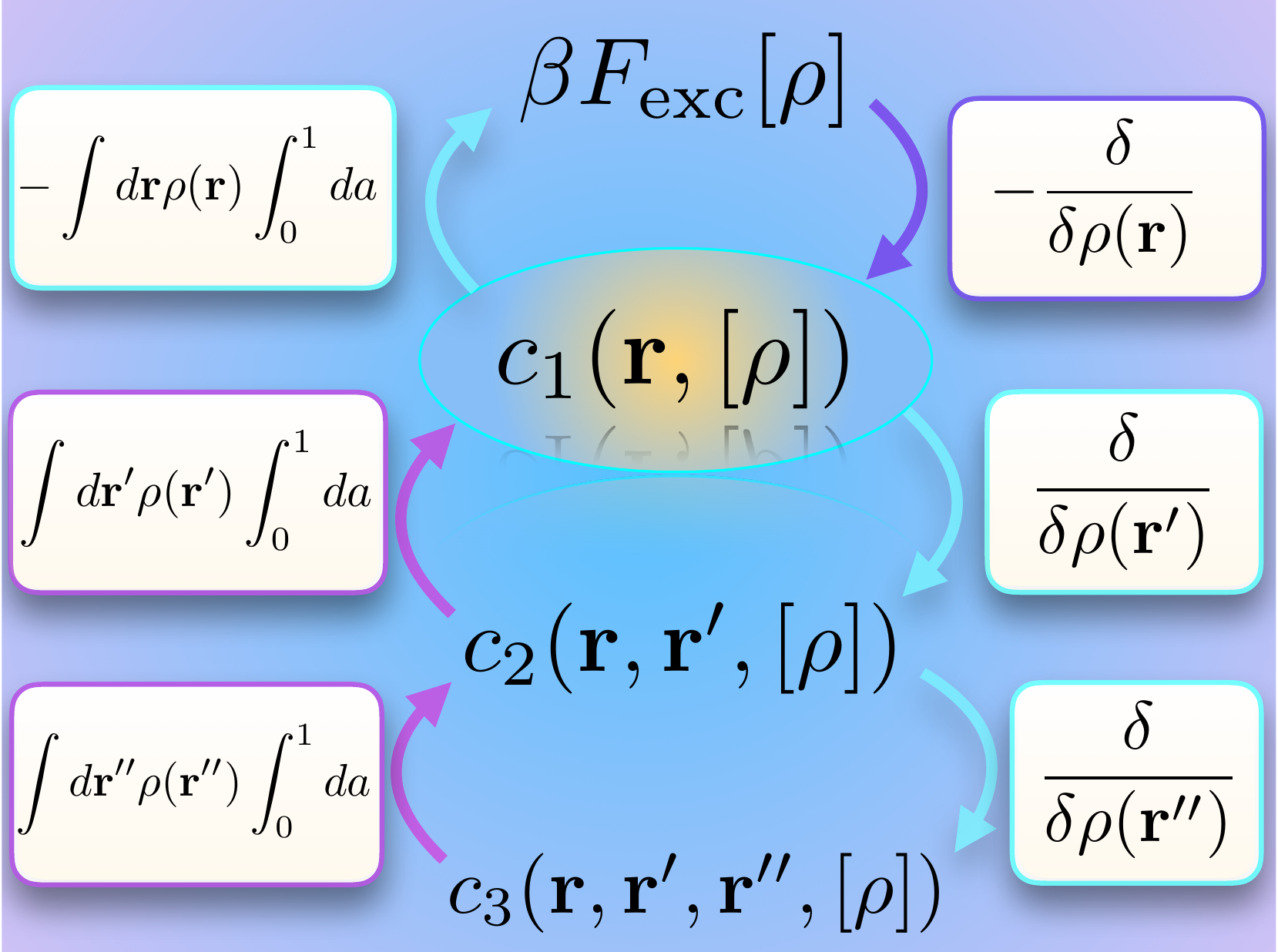}
\caption{Illustration of four different generations of density
  functionals.  Shown are the excess free energy functional
  $F_\rmexc[\rho]$ and the one-, two-, and three-body direct
  correlation functionals.  Upward arrows indicate the relationship
  via functional integration $\int d\rv\rho(\rv) \int_0^1 da$ with the
  integrand being evaluated at the scaled density
  $a\rho(\rv)$. Downward arrows indicate functional differentiation
  $\delta/\delta\rho(\rv)$. The neural functional theory is based on
  training $c_1(\rv,[\rho])$ as the generating mother functional.
  Implementing the arrowed operations only requires high-level
  code. The resulting neural networks, as well as functionals derived
  from analytical expressions, are highly performant.}
\label{FIGfamily}
\end{figure}

While we have emphasized above the role of the one-body direct
correlation functional $c_1(\rv,[\rho])$, primarily due to $c_1(\rv)$
being directly measurable via \eqr{EQc1AsRatio}, one typically rather
starts with a parent functional, the excess free energy functional
$F_\rmexc[\rho]$, in standard accounts of classical density functional
theory. The relationship of $F_\rmexc[\rho]$ and $c_1(\rv,[\rho])$ is
established via functional calculus. Functional differentiation, see
Ref.~\cite{schmidt2022rmp} for a practitioner's account, yields
additional position dependence and we use the notation
$\delta/\delta\rho(\rv)$ to denote the {\it functional} derivative
with respect to the function $\rho(\rv)$. Functional integration is
the inverse operation.  We give a brief description of the functional
relationships in the following. An overview is illustrated in
Fig.~\ref{FIGfamily} and we will return for a broader account below in
Sec.~\ref{SECneuralCalculus}.

The method of {\it automatic differentiation}
\cite{baydin2018autodiff} is an integral part of the new computing
paradigm of differentiable programming \cite{chollet2017}. Automatic
differentiation is based on a powerful set of techniques and it
differs from both symbolic differentiation, as facilitated by computer
algebra systems, and from numerical differentiation via finite
difference, as is computational bread and butter. As shown in the
tutorial \cite{sammueller2023neuralTutorial} only high-level code is
required to invoke automatic differentiation, and both neural {\it
  and} analytical functionals can be differentiated with little
effort. As the derivative (of the functional) is with respect to its
entire input data, the method constitutes a representation of a
genuine functional derivative.

We give an overview. In the present context the functional calculus
that relates the one-body direct correlations to the parent excess
free energy functional is given by the following functional
integration and functional differentiation relations:
\begin{align}
  \beta F_\rmexc[\rho] &=
   -\int d\rv \rho(\rv) \int_0^1 da c_1(\rv,[\rho_a]),
  \label{EQFexcFromFunctionalIntegral}
  \\
  c_1(\rv,[\rho]) &= 
  -\frac{\delta\beta F_\rmexc[\rho]}{\delta\rho(\rv)}.
  \label{EQc1definition}
\end{align}
In \eqr{EQFexcFromFunctionalIntegral} we have parameterized the
general formal integral $\int D[\rho]$ by using $\rho_a(\rv)$ as a
scaled version of the density profile, with a simple linear
relationship $\rho_a(\rv)=a\rho(\rv)$. Hence the parameter value $a=0$
corresponds to vanishing density and $a=1$ reproduces the target
density profile, as it occurs in the argument of $\beta
F_\rmexc[\rho]$ on the left hand side of
\eqr{EQFexcFromFunctionalIntegral}. We emphasize that the integral
over $a$ in \eqr{EQFexcFromFunctionalIntegral} is a simple
one-dimensional integral over the coupling parameter $a$.  The
consistency between Eqs.~\eqref{EQFexcFromFunctionalIntegral} and
\eqref{EQc1definition} is demonstrated below in
Sec.~\ref{SECfunctionalIntegralSumRules}.

The perhaps seemingly very formal functional calculus acquires new and
pressing relevance in light of the neural functional concepts of
Ref.~\cite{sammueller2023neural}, which allows to work explicitly with
both functional derivatives and functional integrals, which can be
evaluated efficiently via the corresponding standalone neural
functionals.

In light of these benefits it is fortunate that the functional
differentiation-integration structure extends recursively to higher
orders of correlation functions.  The next level beyond
Eqs.~\eqref{EQFexcFromFunctionalIntegral} and \eqref{EQc1definition}
involves the two-body direct correlation functional
$c_2(\rv,\rv',[\rho])$ and the integration and differentiation
structure is as follows:
\begin{align}
  c_1(\rv,[\rho]) &=
  \int d\rv' \rho(\rv') \int_0^1 da  c_2(\rv,\rv',[\rho_a]),
  \label{EQc1AsFunctionalIntegral}
  \\
  c_2(\rv,\rv',[\rho]) &= 
  \frac{\delta c_1(\rv,[\rho])}{\delta\rho(\rv')},
  \label{EQc2AsFunctionalDerivative}
\end{align}
and we refer to Refs.~\cite{hansen2013, schmidt2022rmp, brader2013noz,
  brader2014noz} for background.

We can chain the functional derivatives together by inserting
$c_1(\rv,[\rho])$ as given by \eqr{EQc1definition} into the
definition~\eqref{EQc2AsFunctionalDerivative} of
$c_2(\rv,\rv',[\rho])$.  In parallel, we can also chain the functional
integrals in Eqs.~\eqref{EQFexcFromFunctionalIntegral} and
\eqref{EQc1AsFunctionalIntegral}. These procedures yield the following
second order functional integration and differentiation relationships:
\begin{align}
  \beta F_\rmexc[\rho] &= 
  -\int d\rv \rho(\rv)  \int d\rv' \rho(\rv') \notag\\&\qquad
  \times \int_0^1 da \int_0^a da' c_2(\rv,\rv',[\rho_{a'}]),
  \label{EQFexcFromTwoFunctionalIntegrals}
  \\
  c_2(\rv,\rv',[\rho]) &= 
  -\frac{\delta^2 \beta F_\rmexc[\rho]}{\delta\rho(\rv)\delta\rho(\rv')},
  \label{EQc2AsSecondFunctionalDerivative}
\end{align}
where the scaled density profile in
Eq.~\eqref{EQFexcFromTwoFunctionalIntegrals} is
$\rho_{a'}(\rv)=a'\rho(\rv)$. The double parameter integral in
\eqr{EQFexcFromTwoFunctionalIntegrals} can be further simplified
\cite{evans1992}, as described at the end of
Sec.~\ref{SECfunctionalIntegralSumRules}.  The generalization of
\eqr{EQc2AsSecondFunctionalDerivative} to the $n$-th functional
derivative defines the $n$-body direct correlation functional, which
remains functionally dependent on the density profile and which
possesses spatial dependence on $n$ position arguments. Although
increasing $n$ yields objects that become very rapidly out of any
practical reach, the neural functional concept provides much fuel for
making progress. While we do not cover $c_3(\rv,\rv',\rv'',[\rho])$
here, Samm\"uller {\it et al.}\ have demonstrated its general
accessiblity and physical validity for bulk fluids in
Ref.~\cite{sammueller2023neural}.

We have so far focused on the properties of the intrinsic excess free
energy functional $F_\rmexc[\rho]$ and its density functional
derivatives.  This is natural as classically $F_\rmexc[\rho]$ is the
central object that contains the effects of the interparticle
interactions and thus depends in a nontrivial way on its input density
profile. The functional $F_\rmexc[\rho]$ is intrinsic in the sense
that it is independent of external influence.  We recall that we here
work in the grand ensemble (see e.g.\ Refs.~\cite{gonzalez1997,
  white2000, dwandaru2011, delasheras2014canonical} for studies
addressing the canonical ensemble of fixed particle number). Hence the
appropriate thermodynamic potential is the grand canonical free energy
or grand potential. This is required in order to determine
$\rho(\rv)$.

When expressed as a density functional the grand potential consists of
the following sum of ideal, excess, external, and chemical potential
contributions:
\begin{align}
  \Omega[\rho] &= F_\rmid[\rho] + F_\rmexc[\rho]
  + \int d\rv\rho(\rv)[V_\rmext(\rv)-\mu].
  \label{EQomegaFunctional}
\end{align}
The form of the ideal gas free energy functional is explicitly known
as $F_\rmid[\rho]=k_BT \int d\rv \rho(\rv)[\ln\rho(\rv)-1]$ and the
third term in \eqr{EQomegaFunctional} contains the effects of the
external potential $V_\rmext(\rv)$ and of the particle bath at
chemical potential $\mu$.

The variational principle of classical density functional theory
\cite{mermin1965, evans1979, dwandaru2011} ascertains that
\begin{align}
  \frac{\delta \Omega[\rho]}{\delta\rho(\rv)}
  \Big|_{\rho=\rho_0}&= 0 \quad {\rm (min)},
  \label{EQomegaMinimal}\\
  \Omega[\rho_0] &= \Omega_0.
  \label{EQomegaZero}
\end{align}
Equations \eqref{EQomegaMinimal} and \eqref{EQomegaZero} imply that
the grand potential becomes minimal at $\rho_0(\rv)$, which is the
real, physically realized density profile and $\Omega_0$ is the
equilibrium value of the grand potential. Recall that based on the
many-body picture we have $\Omega_0 = -k_BT\ln\Xi$ with the grand
ensemble partition sum $\Xi=\Tr {\rm e}^{-\beta(H-\mu N)}$. We have
used the subscript 0 to denote equilibrium but we drop this elsewhere
in our presentation to simplify notation.

Inserting \eqr{EQomegaFunctional} into \eqr{EQomegaMinimal} and using
the explicit form of the ideal free energy functional together with
the definition \eqr{EQc1definition} of $c_1(\rv,[\rho])$ leads to
\eqr{EQel} with the one-body direct correlations expressed as a
density functional, as anticipated in
Sec.~\ref{SECdft}. Exponentiating and regrouping the terms then yields
the following popular form of the Euler-Lagrange equation:
\begin{align}
  \rho(\rv) = 
  \exp\Big( -\beta V_\rmext(\rv) + \beta \mu + c_1(\rv,[\rho]) \Big).
  \label{EQelExpoentiated}
\end{align}
Equation \eqref{EQelExpoentiated} is a self-consistency relation that
can be solved efficiently for the equilibrium density profile
$\rho(\rv)$ via iterative methods, as detailed below in
Sec.~\ref{SECapplication}.  A~prerequisite is that $c_1(\rv,[\rho])$
is known, usually as an approximation that is obtained from an
approximate excess free energy functional $F_\rmexc[\rho]$ via
functionally differentiating according to \eqr{EQc1definition}. Having
obtained a numerical solution of \eqr{EQelExpoentiated} for the
density profile, this can then be inserted into the grand potential
functional \eqref{EQomegaFunctional} to obtain full thermodynamic
information via \eqr{EQomegaZero}, which by construction is consistent
with the density profile.

We demonstrate in the following how this classical functional
background can be put to formidable use via hybridization with
simulation-based machine learning. As our aim is pedagogical, we
choose the one-dimensional hard core system as a concrete example to
demonstrate the general methodology \cite{sammueller2023neural}. We
complement the neural functional structure with a description of
Percus' analytical solution, which then allows for mirroring of the
neural theory.

\section{Neural functional theory}
\label{SECnft}

\begin{figure}[!t]
  \includegraphics[page=1,width=.99\columnwidth]{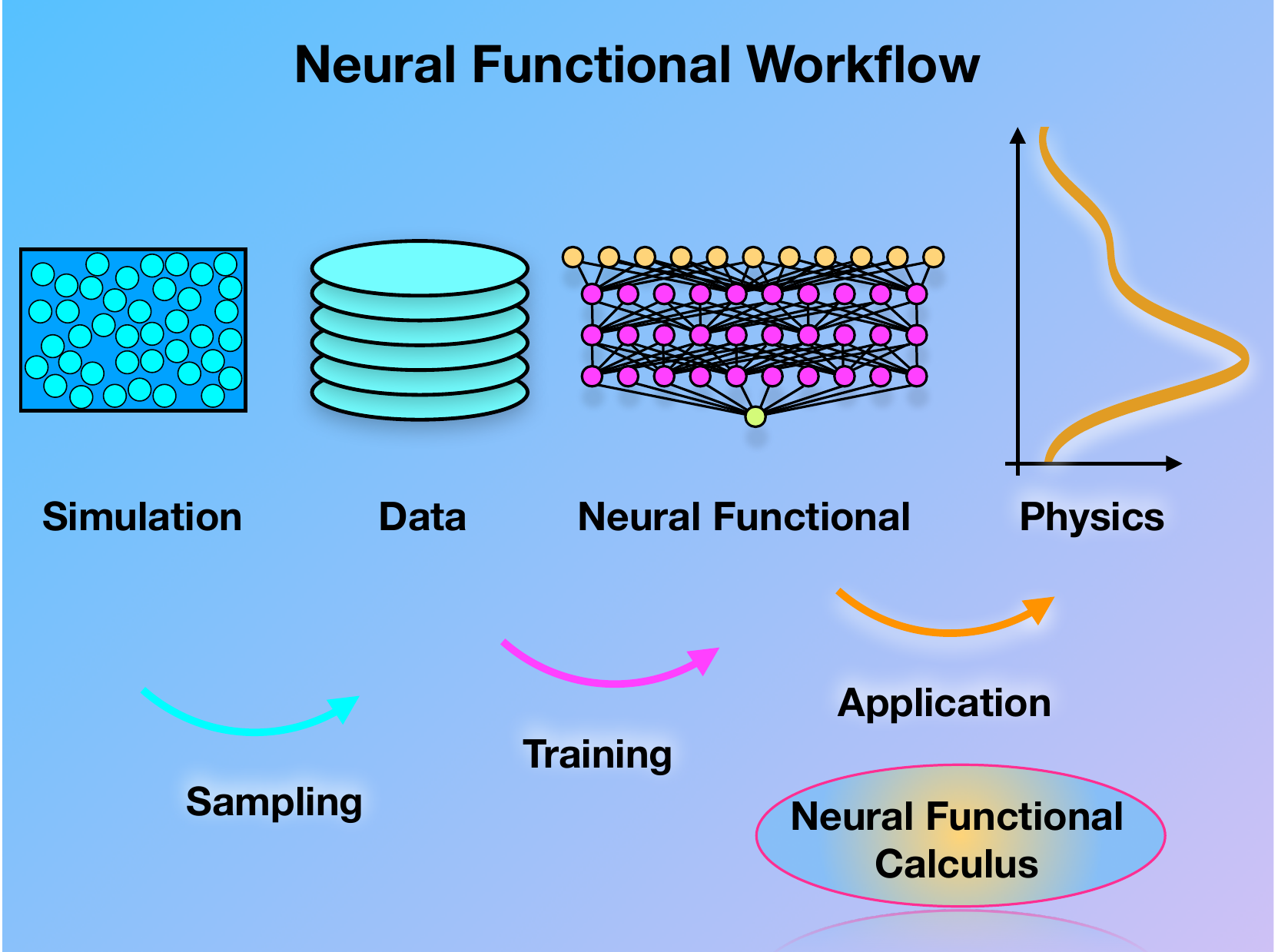}
\caption{Schematic of the workflow of the neural functional
  theory. Many-body simulations under randomized conditions are used
  to sample statistically averaged and spatially resolved data that
  characterize the inhomogeneous response of the considered system.  A
  neural network is then trained to represent the direct correlation
  functional, which is subsequently applied numerically and via neural
  functional methods to investigate the physics of the system in the
  desired target situations.  }
\label{FIGworkflow}
\end{figure}

Jerry Percus famously wrote in the abstract of his 1976 statistical
mechanics landmark paper \cite{percus1976}: ``The external field
required to produce a given density pattern is obtained explicitly for
a classical fluid of hard rods. All direct correlation functions are
shown to be of finite range in all pairs of variables.'' Here we
relate his achievement to the neural functional theory, which allows
to reproduce numerically a variety of properties of the exact
solution.  We emphasize that the neural functional theory remains
generic in its applicability to further model fluids; see the
Supplementary Information of Ref.~\cite{sammueller2023neural} for the
successful treatment of the supercritical Lennard-Jones fluid in three
dimensions.  We refer the Reader to the provided online resources
\cite{sammueller2023neuralTutorial} for a programming tutorial on the
concrete application of the following concepts. Figure
\ref{FIGworkflow} shows a schematic of the workflow that is inherent
in the neural functional concept, as described in the following.

\subsection{Training the neural correlation functional}
\label{SECtraining}

The classical fluid of hard rods that Percus considers has
one-dimensional position coordinates $x_i$, with particle index
$i=1,\ldots,N$ and a pairwise interparticle interaction potential
$\phi(x)$ which is infinite if the distance $x$ between the two
particles is smaller than their diameter, $x<\sigma$, and it vanishes
otherwise. The system is exposed to an external potential
$V_\rmext(x)$, which is a function of position $x$ across the system,
and this in general creates an inhomogeneous ``density pattern''
$\rho(x)$.

We adjust the definition \eqref{EQrhoAverage} of the density
distribution to the present one-dimensional case:
\begin{align}
  \rho(x) = \avg{\sum_i\delta(x-x_i)},
  \label{EQrhoCountingOneDimension}
\end{align}
where $\delta(\cdot)$ here indicates the Dirac distribution in one
dimension and the brackets indicate a grand canonical thermal
average. Due to the hard core nature of the model, the statistical
weight of each ``allowed'' microstate is particularly simple and given
by $\exp[-\beta\sum_i V_\rmext(x_i)+\beta\mu N]/\Xi$, where $\Xi$ is a
normalizing factor.  Allowed microstates are those for which all
distinct particle pairs $ij$ are spaced far enough apart, $|x_i-x_j|
\geq \sigma$. If already a single overlap occurs, then the microstate
is ``forbidden'' as the interparticle potential becomes formally
infinite, which then creates vanishing statistical weight; we recall
the illustration in Fig.~\ref{FIGrods}.

Despite the apparent simplicity of the many-body probability
distribution, the Statistical Mechanics of the hard rod model is
nontrivial. The particles interact nonlocally over the lengthscale
$\sigma$ and the external potential has no restrictions on its shape
or on the lengthscale(s) of variation. Hence features such as jumps
and positive infinities that represent hard walls are allowed.  In
bulk, $V_\rmext(x)=0$, and the solution is straightforward
\cite{tonks1936, hansen2013}.  The general case is however highly
nontrivial, which makes Percus' above quoted opening a very remarkable
one. We present more details of his work further below, after first
laying out the general machine learning strategy of
Ref.~\cite{sammueller2023neural}. This neural functional method is
neither restricted to hard cores nor to one-dimensional systems, but
addressing this case here is useful to highlight the salient features
of the approach.

We aim for explicitly sampling the microstates of the system according
to their probability distribution via particle-based simulations. This
can be implemented efficiently, and for the present introductory
purposes in also an intuitively accessible way, via grand canonical
Monte Carlo (GCMC) sampling.  Excellent accounts of this method are
given in Refs.~\cite{frenkel2023book, hansen2013, wilding2001,
  brukhno2021dlmonte}. Briefly, a Markov chain of microstates is
constructed, where based on a given configuration, a trial step is
proposed, which is accepted with a probability given by the Metropolis
function $\min[1,\exp(-\beta\Delta E)]$, where $\Delta E$ is the
energy difference between the original and the trial state.

Three trial moves are used in the simplest yet powerful scheme: i)
Selecting one particle $i$ randomly and displacing it uniformly within
a given maximal cutoff distance. If the displacement creates overlap,
then the trial move is discarded. If otherwise there is no overlap in
the new configuration, the energy difference is due to only the
external potential, $\Delta E=V_\rmext(x_i)-V_\rmext(x_i')$, where the
prime denotes the trial position of particle $i$. ii) A new particle
$j$ is inserted at a random position $x_j$ with energy change that
accounts for both the external potential and the chemical equilibrium
with the particle bath and hence $\Delta E = V_\rmext(x_j')-\mu$. iii)
Correspondingly, a randomly selected particle $i$ is removed from the
system. The acceptance of the removal happens again with a probability
given by the Metropolis function with energy difference $\Delta E= -
V_\rmext(x_i) + \mu$.

Despite its conceptual simplicity GCMC is a very powerful method for
the investigation of complex effects \cite{frenkel2023book,
  wilding2001, brukhno2021dlmonte} and significant extensions exist
both in the form of histogram techniques \cite{wilding2001,
  brukhno2021dlmonte} and the tailoring of more complex and collective
trial moves.  Investigating a typical physical problem, as specified
by the interparticle interactions $u(\rv^N)$ and the type of
considered external influence, such as walls as represented by a model
form of $V_\rmext(\rv)$, requires e.g.\ scanning of the thermodynamic
parameters and acquiring good enough statistics at each
statepoint. Our ultimate goal (Sec.~\ref{SECapplication}) is to
perform this tasks with significant gain in efficiency via the neural
theory; we re-iterate the availability via
Ref.~\cite{sammueller2023neuralTutorial} of hands-on code examples for
the present hard rod model.

We base the training on the following rewriting and adaption of the
chemical potential balance \eqr{EQel} to the one-dimensional system:
\begin{align}
  c_1(x) &= \ln \rho(x) + \beta V_\rmext(x) - \beta \mu.
  \label{EQelc1}
\end{align}
All quantities on the right hand side are either prescribed a priori
or are accessible via the GCMC simulations: Specifically, the density
profile $\rho(x)$ is obtained by filling a position-resolved histogram
according to the encountered microstates as specified by its particle
coordinates~$x_i$. We recall the formal definition
\eqref{EQrhoCountingOneDimension} of $\rho(x)$ via the Dirac
distribution, which in practice is discretized such that sufficient
finite spatial resolution, say $0.01\sigma$, is obtained. This
``counting'' method is arguably the most intuitive one to obtain data
for the density profile.  As an aside, there is a number of
force-sampling techniques that can improve the statistical variance
significantly \cite{borgis2013, delasheras2018forceSampling,
  rotenberg2020, robitschko2023any} and that also can serve to gauge
the quality of sampling of the equilibrium ensemble
\cite{robitschko2023any}.

While the issues of Monte Carlo sampling efficiency and quality
assessment of thermal averages can be pertinent in higher dimensions
and in physically more complex situations, the simplicity of the
present one-dimensional hard core model makes counting according to
\eqr{EQrhoCountingOneDimension} an appropriate choice to obtain data
for $\rho(x)$. Then adding up the three contributions on the right
hand side of \eqr{EQelc1} yields results for $c_1(x)$.  We proceed at
this data-generation stage somewhat heretically and ignore at first
the central role that $c_1(x)$ plays for the physics of inhomogeneous
systems.

\begin{figure}[!t]
  \includegraphics[page=1,width=.99\columnwidth]{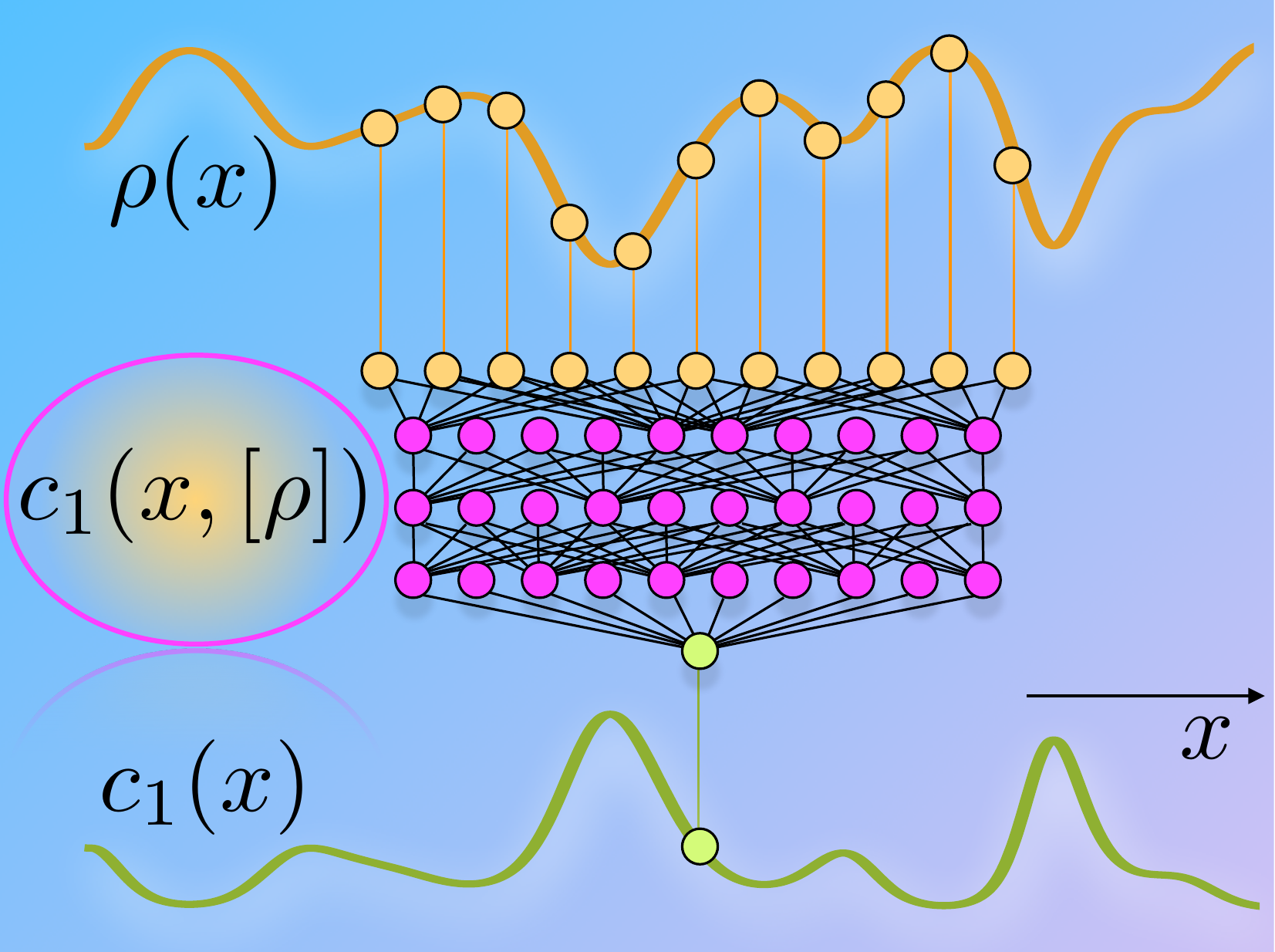}
\caption{Illustration of the neural one-body direct correlation
  functional $c_1^\star(x,[\rho])$ represented by a fully connected
  neural network with three hidden layers. The topology maps a small
  finite window of the density profile $\rho(x)$ to the local value of
  the direct correlation function $c_1(x)$.}
\label{FIGnetwork}
\end{figure}

In contrast to the typical deterministic setup for investigating a
specific physical situation described above, training the neural
network proceeds on the basis of randomized situations rather than
with the ultimate application in mind; we recall the illustration of
the neural functional workflow shown in Fig.~\ref{FIGworkflow}. The
motivation for using this strategy comes from the goal of capturing
via the machine learning the intrinsic direct correlations of the
many-body system that then transcend the specific inhomogeneous
situations that were under consideration during training. Figure
\ref{FIGnetwork} depicts an illustration of the neural network
topology of the trained central neural network $c_1^\star(x,[\rho])$
and its relation to the physical input and output quantities, i.e., to
$\rho(x)$ and $c_1(x)$.

We hence perform a sequence of simulation runs, where each run has an
input value $\beta \mu^{(k)}$ and an input functional shape $\beta
V_\rmext^{(k)}(x)$, both of which are generated randomly.
Specifically, we combine sinusoidal functions with periodicities that
are commensurate with the box length~$L$, linear discontinuous
segments, and hard walls in the creation of $V_\rmext^{(k)}(x)$; see
Refs.~\cite{sammueller2023neural, sammueller2023neuralTutorial} for
further details.  The superscript $k$ enumerates the different GCMC
simulation runs and in practice we perform 512 of these. The result is
a set of corresponding density profiles $\rho^{(k)}(x)$. We then use
\eqr{EQelc1} to obtain for each run the one-body direct correlation
profiles from simply adding up: $ c_1^{(k)}(x) = \ln \rho^{(k)}(x) +
\beta V_\rmext^{(k)}(x) - \beta \mu^{(k)}$. As a result of the
simulation protocol we have generated a bare data set $\{
\beta\mu^{(k)}, \beta V_\rmext^{(k)}(x)$, $\rho^{(k)}(x)$,
$c_1^{(k)}(x) \}$ for all positions $x$ and for all different runs
$k$. {\ms As a practical detail, this requires to exclude regions
  where $\rho(x)=0$ and $V_\rmext(x)=\infty$.}

In order to address our declared goal to learn a functional dependence
of $c_1(x)$, we have to carve out a nontrivial dependence relationship
and hence restrict the data input. Motivated by the physics, one might
see the scaled chemical potential $\beta \mu^{(k)}$ and the scaled
external potential $\beta V_\rmext^{(k)}(x)$ to be the true mechanical
origin of the shape of the direct correlation function
$c_1^{(k)}(x)$. However, the insights provided by density functional
theory hint at the fact that this is not the best possible choice of
functional relationship to consider.

We recap that the GCMC simulations yield data according to:
\begin{align}
  \big\{V_\rmext^{(k)}(x')- \mu^{(k)} \big\}_0^L 
  &\;\longrightarrow\; 
  \big\{\rho^{(k)}(x)\big\}_0^L,
\end{align}
where the curly brackets indicate all function values inside of the
system box, with ranges $0 \leq x' \leq L$ and $0\leq x \leq L$; the
arrow indicates an input-output relationship. Applying \eqr{EQelc1} to
the entire data set also allows to have the direct correlation
function as an output according to:
\begin{align}
  \big\{
   V_\rmext^{(k)}(x') - \mu^{(k)}
  \big\}_0^L
  &\;\longrightarrow\;
  \big\{
   c_1^{(k)}(x)
  \big\}_0^L.
  \label{EQmockSimulation}
\end{align}
If one were to mimic the simulations directly by the neural network
one would be tempted to base the training directly upon
\eqr{EQmockSimulation}. In less clearcut machine-learning situations
than considered here, it can be a standard strategy to attempt to
represent the causal relationship, which governs the complex
mathematical or real-world system under consideration, by a surrogate
artificial intelligence model. The present functional formulation of
Statistical Mechanics hints at potential caveats, such as the
necessity of dealing with the full input and output data sets
(parameter ranges of $x$ and $x'$) across the entire
system. Furthermore the specific physics of the mutually interacting
rods appears to play no role.

The density functional-inspired training (Sec.~\ref{SECdft}) proceeds
very differently. We here take a pragmatic stance and attempt to
create via training a neural representation of the dependence of
$c_1(x)$ on $\rho(x)$ alone. This leads to a surrogate model
$c_1^\star(x,[\rho])$ based on the following mapping
\begin{align}
  \big\{\rho^{(k)}(x')\big\}_{x-x_c}^{x+x_c}
  \;\longrightarrow\;
  c_1^{(k)}(x),
  \label{EQrhoToc1inTraining}
\end{align}
where the input on the left hand side consists of function values
$\rho^{(k)}(x')$ that lie inside the density window centered at $x$,
i.e., only the values $x'$ that lie within a narrow interval $x-x_c
\leq x' \leq x+x_c$.  Here $x_c$ is a cutoff parameter that for
short-ranged interparticle potentials is of the order of the particle
size. For the present one-dimensional hard core system we set
$x_c=\sigma$. Instead of having to output an entire function, as would
be the case when attempting to learn via \eqr{EQmockSimulation}, here
the output is merely the single value of the direct correlation
function at the center of the density window.  We recall that this
target value is obtained from the simulation data via \eqr{EQelc1}
such that $c_1^{(k)}(x)= \ln\rho^{(k)}(x)+\beta
V_\rmext^{(k)}(x)-\beta \mu^{(k)}$ for each run $k$.  A simple GCMC
code is provided online \cite{sammueller2023neuralTutorial}, along
with a pre-generated simulation data set and a pre-trained neural
functional.

{\ms We choose the loss function to be the mean squared error of the
  neural network output compared to the simulation reference value for
  $c_1(x)$, as obtained via \eqr{EQelc1}. As a further metric to gauge
  the training progress, we make use of the mean absolute error of
  reference and output. Both choices are standard
  \cite{chollet2017}. The quadratic loss is convenient as it is
  analytical and hence the machine-learning gradient-based methods
  directly apply. The mean absolute error is nonanlytical due to the
  modulus involved, but it is a useful supporting quantity that has a
  very direct interpretation.

  After training the mean absolute error was of the order of $\sim
  0.013$, which implies that the neural network prediction deviates on
  average by this value from the simulation data. Although the
  simulation data carries some statistical noise, its effect is
  comparatively smaller, when taking the numerical solution of the
  Percus theory (detailed below) as the reference.

  Our training data consists of 512 simulation runs using a simulation
  box size of $L=10\sigma$. Each of the simulation runs requires only
  about three minutes runtime on a single CPU core of a standard
  desktop machine.

  We use a standard fully-connected artificial neural network with
  three hidden layers that respectively possess 128, 64 and 32 nodes.
  We use 201 input nodes to represent the density profile in a finite
  window of size $1\sigma$ and spatial bin size $0.01 \sigma$, where
  we recall that $\sigma$ is the particle size.  To accommodate the
  local functional mapping, we reshape the training data into input
  density windows and corresponding output values of $c_1(x)$, where
  we also apply twofold data augmentation by exploiting mirror
  symmetry of the simulation results. Excluding regions where
  $V_\rmext(x) = \infty$ and hence where \eqr{EQelc1} is not defined,
  this results in $\sim 10^6$ input-output pairs.

}

From the above description and without considering the background in
density functional theory it is not evident that the training will be
successful and minimize the loss satisfactorily to yield a trained
network $c_1^\star(x,[\rho])$. From a mathematical point of view, this
raises the questions whether a corresponding object $c_1(x,[\rho])$
indeed exist and whether it is unique.  And if so, is its structure
simple enough that it can be written down explicitly?

\subsection{Percus' exact direct correlation functional}
\label{SECpercusFunctional}
Due to Percus singular achievement \cite{percus1976} the one-body
direct correlation functional $c_1(x,[\rho])$ for interacting hard
rods in one spatial dimension is known analytically and this has
triggered much subsequent progress, see e.g.\ Refs.~\cite{robledo1981,
  vanderlick1989, bakhti2012, rosenfeld1988, rosenfeld1989,
  tarazona2000, roth2002WhiteBear, roth2006WhiteBear, roth2010}. The
functional dependence on the density profile is nonlocal, as one would
expect from the fact that the rods interact over the finite distance
$\sigma$, and it is also nonlinear, as is consistent with the
behaviour of a nontrivially interacting many-body system. The spatial
dependence is characterized by convolution operations which, despite
performing the task of coarse-graining, retain the full character of
the microscopic interactions. The Percus functional provided
motivation for developing so-called weighted-density approximations
(WDA) \cite{hansen2013}, where the density profile is convolved with
one or several weight functions that are then further processed to
give the ultimate value of the density functional.

We here give the Percus direct correlation functional in Rosenfeld's
geometry-based fundamental measure representation, see
Ref.~\cite{percus2013} for a historical perspective. Instead of
working with the particle diameter~$\sigma$ as the fundamental
lengthscale, Rosenfeld rather bases his description on the particle
radius $R=\sigma/2$, which allows to find deep geometric meaning in
Percus' expressions and to also generalize to higher dimensions
\cite{rosenfeld1988, rosenfeld1989, roth2010}.

The exact form \cite{rosenfeld1988} of the one-body direct correlation
functional is analytically given as the following sum:
\begin{align}
  c_1(x,[\rho]) 
   = & - \frac{\Phi_0(x-R)+ \Phi_0(x+R)}{2} 
  \notag \\&
  - \int_{x-R}^{x+R} dx' \Phi_1(x').
  \label{EQc1Percus}
\end{align}
Here the two functions $\Phi_0(x)$ and $\Phi_1(x)$ each depend on two
weighted densities $n_0(x)$ and $n_1(x)$ in the following form:
\begin{align}
  \Phi_0(x) &= -\ln[1-n_1(x)],
  \label{EQphi0}\\
  \Phi_1(x) &= \frac{n_0(x)}{1-n_1(x)}.
  \label{EQphi1}
\end{align}
The weighted densities $n_0(x)$ and $n_1(x)$ are obtained from the
bare density profile via spatial averaging:
\begin{align}
  n_0(x) &= \frac{\rho(x-R) + \rho(x+R)}{2},
  \label{EQn0definition}\\
  n_1(x) &= \int_{x-R}^{x+R} dx' \rho(x').
  \label{EQn1definition}
\end{align}
The discrete spatial averaging at positions $x\pm R$ in the weighted
density \eqref{EQn0definition} parallels that in the first term of
\eqr{EQc1Percus}. Similarly the position integral over the interval
$[x-R,x+R]$ in \eqr{EQn1definition} appears analogously in the second
term of \eqr{EQc1Percus}. These similarities are not by coincidence.
The structure is rather inherited from the grandmother (excess free
energy) functional, as is described in Sec.~\ref{SECintegrating}.

Having the analytical solution
\eqref{EQc1Percus}--\eqref{EQn1definition} for $c_1(x,[\rho])$ allows
for carrying out numerical evaluation and comparing against results
from the neural functional $c_1^\star(x,[\rho])$. The range of
nonlocality, i.e., the distance across which information of the
density profile enters the determination of $c_1(x,[\rho])$ via
Eqs.~\eqref{EQc1Percus}--\eqref{EQn1definition} is strictly finite, as
announced in Percus' abstract \cite{percus1976}. As two averaging
operations, each with range $\pm R$, are chained together, the
composite procedure has a range of $\pm 2R=\pm\sigma$, inline with our
truncation of the density profiles in the training data sets according
to \eqr{EQrhoToc1inTraining}.  A numerical implementation of Percus
direct correlation functional is available online
\cite{sammueller2023neuralTutorial}.

\subsection{Application inside and beyond the box}
\label{SECapplication}

\begin{figure*}[htb!]
  \includegraphics[width=.99\linewidth]{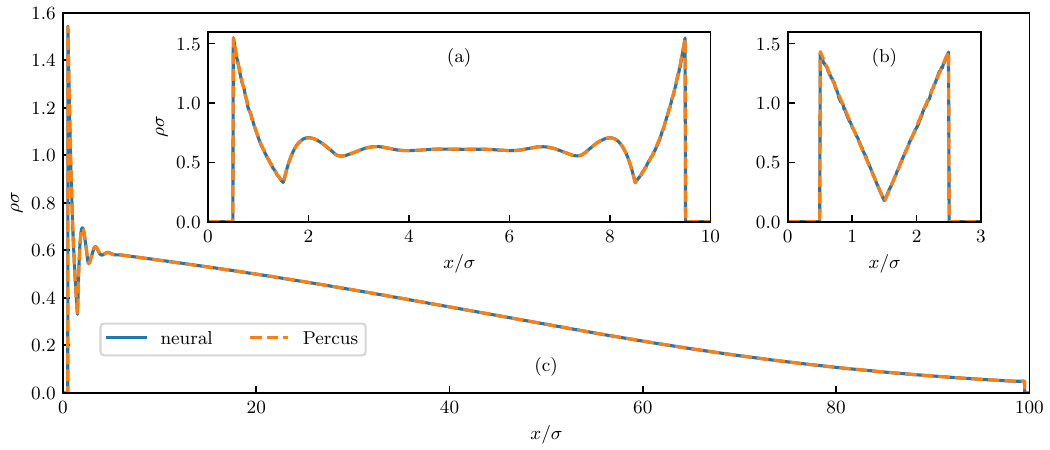}
\caption{Representative density profiles that the inhomogeneous
  hard rod system exhibits under the influence of an external
  potential. The results are obtained from numerically solving
  \eqr{EQelOneDimension} upon using either the neural direct
  correlation functional $c_1(x,[\rho])$ or Percus exact solution
  therof. The three cases comprise (a) two hards wall with separation
  distance $9\sigma$ and chemical potential $\beta \mu=2$, (b) two
  hard walls with much smaller separation distance $2\sigma$ and
  identical chemical potential $\beta\mu=2$, and (c)
  sedimentation-diffusion equilibrium under gravity with a locally
  varying chemical potential, $\beta\mu_{\rm loc}(x) = \beta\mu-\beta
  V_\rmext(x) = 2-0.05x/\sigma$; here the linearly varying
  contribution accounts for the influence of gravity on the system and
  confinement is provided by two widely spaced hard walls at
  $x=0.5\sigma$ und $x=99.5\sigma$.  Note the crossover in panel (c)
  from the strongly oscillatory behaviour near the lower wall to a
  very smooth density decay, effectively following a local density
  approximation \cite{hansen2013}, upon increasing the scaled height
  $x/\sigma$.
  \label{FIGrhoProfiles}
  }
\end{figure*}

Actually making the predictions for the hard rod model is now
straightforward as we can resort to density functional theory and its
standard use in application to physical problems. The arguably most
common method for solving the Euler-Lagrange equation
self-consistently is based on \eqr{EQelExpoentiated}, which we
re-express for the one-dimensional case considered:
\begin{align}
  \rho(x) = \exp\Big(
  -\beta V_\rmext(x) + \beta \mu + c_1^\star(x,[\rho])
  \Big).
  \label{EQelOneDimension}
\end{align}

We recall that the range of nonlocality of $c_1(x,[\rho])$ is limited
to only the particle size $\sigma$ and that we were able to extract
the functional dependence from simulation data obtained by sampling in
boxes of size $L$. Although the value of $L$ could in principle be
imprinted in subtle finite size effects that $c_1^\star(x,[\rho])$ has
acquired, the size $L$ of the original simulation box has vanished and
the application of the neural functional in \eqr{EQelOneDimension} is
fit for use to predict properties of much larger systems. As an
example, Ref.~\cite{sammueller2023neural} demonstrates the scaling up
by a factor of 100 from the original simulation box to the predicted
system of three-dimensional hard spheres under gravity.

The numerical solution of \eqr{EQelOneDimension} can be efficiently
performed on the basis of Picard iteration where an initial guess of
the density profile is inserted on the right hand side and the
resulting left hand side is used to nudge the initial guess in the
correct direction toward the self consistent solution. This is
numerically fast and straightforward to implement, see the
tutorial~\cite{sammueller2023neuralTutorial}. A common choice is to
mix five percent of the new solution to the prior estimate.

{\ms 

As laid out above, we choose the one-dimensional hard core model due
to both the availability of Percus' functional and the computational
ease of both numerical evaluation of the analytical expressions and of
carrying out many-body simulations. On the downside, the model does
not form a very credible platform for assessing the numerical
efficiency gain of the neural theory, as in general one will be
interested in more complex systems and more complex physical
situations than addressed here. Nevertheless, to give a rough idea
about the required computational workload, minimizing the neural
density functional takes of the order of seconds on a GPU, while the
GCMC simulation runtime is of the order of several minutes. Minimizing
the analytical Percus functional is faster than using the neural
network, due to the simple structure of
Eqs.~\eqref{EQc1Percus}--\eqref{EQn1definition}, which facilitates
using very high-performance fast Fourier transforms.

}

We show three representative examples of density profiles for narrow
to wide confinement between impenetrable walls in
Fig.~\ref{FIGrhoProfiles}. In all cases the results from using the
neural functional are numerically identical to those from the Percus
functional on the scale of the plot. The profiles in narrow
[Fig.~\ref{FIGrhoProfiles}(a)] and in moderately wide
[Fig.~\ref{FIGrhoProfiles}(b)] pores show very dinstinct features with
the strongly confined system in (a) having a striking V-shape, which
arises from having at most two particles in the system, to the more
generic damped oscillatory behaviour in the moderately wide pore (b).
The main panel Fig.~\ref{FIGrhoProfiles}(c) shows the influence of a
weak gravitational field, which creates a continuously varying density
inhomogeneity across the entire system. The decay in local density
occurs with a much larger length scale as compared to the particle
packing effects that are localized near the lower wall.

The behaviour shown in Fig.~\ref{FIGrhoProfiles}(c) away from the
walls is well-represented by a local density approximation
\cite{hansen2013} (see e.g.\ Ref.~\cite{jex2023} for recent
mathematical work).  The local density approximation can be a useful
tool when investigating e.g.\ macroscopic ordering under gravity,
where the occurring stacking sequences of different thermodynamic
phases can be traced back to the phase diagram
\cite{delasheras2015stacking, eckert2021communPhys}. In particular the
effects on mixtures were rationalized by a range of techniques, from
generalization of Archimedes' principle \cite{piazza2008sedimentation,
  piazza2012archimedes} to analyzing stacking sequences
\cite{delasheras2015stacking, eckert2021communPhys}.
{\ms We stress that the present model applications constitute very
  significant extrapolations from the training data that we recall was
  obtained in a fixed box size $L=10\sigma$ and under the influence of
  randomized external and chemical potentials. Hence both the very
  confined system [Fig.~\ref{FIGrhoProfiles}(b)] and the large system
  [Fig.~\ref{FIGrhoProfiles}(c)] constitute significant extrapolations
  from the training data.

}

As a further potential application of the neural functional theory,
the dynamical density functional theory \cite{evans1979, marconi1999,
  archer2004} is similarly easy to implement numerically as
\eqr{EQelOneDimension} and it is a currently popular choice to study
time-dependent problems
\cite{tevrugt2020review,tevrugt2022perspective}. We comment on the
status of the approach \cite{delasheras2023perspective} and how
machine learning can help to overcome its limitations in
Sec.~\ref{SECnonequilibrium} below.

\section{Neural functional calculus}
\label{SECneuralCalculus}

We have seen in Sec.~\ref{SECnft} how a neural one-body direct
correlation functional can be efficiently trained on the basis of a
pool of pre-generated Monte Carlo simulation data that are obtained
under randomized conditions. The specific way of organizing the
simulation data into training sets mirrors the functional
relationships given by classical density functional theory.  We have
then shown that the neural functional can efficiently be used to
address physical problems, taking the one-dimensional hard rod system
as a simple example of a mutually interacting many-body system.

We here proceed by exemplifying the depth of physical insight that can
be explored by acknowledging the functional character of the trained
neural correlation functional. Hence we lay out functional integration
(Sec.~\ref{SECintegrating}) and functional differentiation
(Sec.~\ref{SECdifferentiation}).  We show sum rule construction via
Noether invariance (Sec.~\ref{SECnoether}), via exchange symmetry
(also Sec.~\ref{SECnoether}), and via functional integration
(Sec.~\ref{SECfunctionalIntegralSumRules}). The presentation in each
subsection is self-contained to a considerable degree and we
illustrate the generality of the methods both by application to the
neural functional as well as by revisiting the analytic Percus theory.

\subsection{Functional integration of direct correlations}
\label{SECintegrating}

Having captured the essence of molecular packing effects, as they
arise from the short-ranged hard core repulsion between the particles,
via the neural functional $c_1^\star(x,[\rho])$, begs for speculation
whether additional and as yet hidden physical structure can be
revealed.  We give two plausibility arguments why one should expect to
be able to postprocess $c_1^\star(x,[\rho])$ in a meaningful way to
retrieve global information.

First, thermodynamics is based on the existence of very few and
well-defined unique and global quantities, such as the entropy, the
internal energy, and the free energy. Carrying out parametric
derivatives, with powerful interrelations given by the Maxwell
relations, enables one to obtain equations of state, susceptibilities
and further measurable global quantities. Our neural direct
correlation functional in contrast is a local object with finite range
of nonlocality. So how does this relate to the global information?

The second argument is more formal. Suppose we prescribe the form of
the density profile and then evaluate the neural functional
$c_1^\star(x,[\rho])$ at each position $x$. This procedure yields a
numerical representation of the corresponding direct correlation
function $c_1(x)$. In the practical numerical representation we have a
set of discrete grid points that represent the function values at
these spatial locations~$x$.  Hence the entire data set forms a
numerical array or numerical vector, indexed by $x$. One can then ask
whether this vector could potentially be the gradient of an
overarching parent object?

The physical and the formal question can both be answered
affirmatively due to the existence of the excess free energy density
functional $F_\rmexc[\rho]$. Its practical route of access, based on
functional integration along a continuous sequence of states (a
``line'') in the space of density functions, is strikingly
straightforward within the neural method. The core of the method is to
evaluate $c_1(x,[\rho_a])$ as described above, but for a range of
scaled versions of the prescribed density profile $\rho_a(x)$ and then
integrating in position to obtain the excess free energy as a global
value, see the functional integral given in
\eqr{EQFexcFromFunctionalIntegral}.

Specifically, we define a scaled version of the density profile as
$\rho_a(x) = a\rho(x)$, such that $a=0$ generates the empty state that
has vanishing density profile, $\rho_{a=0}(x)=0$. On the other end
$a=1$ yields the actual density profile of interest,
$\rho_{a=1}(c)=\rho(x)$.  The excess free energy functional is then
obtained easily via functional integration according to
\begin{align}
  \beta F^\star_\rmexc[\rho] &=
  -\int dx \rho(x) \int_0^1 da c_1^\star(x,[\rho_a]).
  \label{EQFexcNeuralFromFunctionalIntegral}
\end{align}
The numerical evaluation requires evaluation of $c_1^\star(x,[\rho_a])$
at all positions $x$ in the system and for a range of intermediate
values $0\leq a \leq 1$ such that the parametric integral over $a$ can
be accurately discretized.

Analytically carrying out the functional integral
\eqref{EQFexcNeuralFromFunctionalIntegral} on the basis of the
analytical direct correlation functional $c_1(x,[\rho])$ as given by
Eqs.~\eqref{EQc1Percus}--\eqref{EQn1definition} is feasible. The
result~\cite{robledo1981}, again expressed in the more illustrative
Rosenfeld fundamental measure form, is given by:
\begin{align}
  \beta F_\rmexc[\rho] &= \int dx \Phi(n_0(x),n_1(x)),
  \label{EQFexcPercusViaPhi}\\
  \Phi(n_1(x),n_2(x)) &= -n_0(x) \ln[1-n_1(x)].
  \label{EQPhiPercus}
\end{align}
Here the integrand $\Phi(n_0(x), n_1(x))$ plays the role of a
localized excess free energy density which depends on the weighted
densities $n_0(x)$ and $n_1(x)$ as given via the spatial averaging
procedures in Eqs.~\eqref{EQn0definition} and \eqref{EQn1definition},
respectively.  Inserting \eqr{EQPhiPercus} into
\eqr{EQFexcPercusViaPhi} yields the hard rod excess free energy
functional in the following more explicit form:
\begin{align}
  \beta F_\rmexc[\rho] &= -\int dx n_0(x)\ln[1-n_1(x)].
  \label{EQFexcPercusExplicit}
\end{align}
Equation \eqref{EQFexcPercusExplicit} is strikingly compact, given
that it describes the essence of a system of mutually interacting hard
cores exposed to an arbitrary external potential. 

Although the result of the functional integral
\eqref{EQFexcNeuralFromFunctionalIntegral} has lost all position
dependence, the specific form of the density profile $\rho(x)$ is
deeply baked into the resulting output value of the functional via
both the prefactor $\rho(x)$ in the integrand in
\eqr{EQFexcNeuralFromFunctionalIntegral} and the evaluation of the
direct correlation functional at the specifically scaled form
$\rho_a(x)$. In parallel with this mathematical structure, the
explicit form \eqref{EQFexcPercusExplicit} of the Percus functional
clearly demonstrates that the resulting value will depend nontrivially
on the shape of the input density profile.

Having demonstrated that $F_\rmexc[\rho]$ as a global quantity can be
obtained from appropriate functional integration of a locally resolved
correlation functional $c_1(x,[\rho])$ naturally leads to the question
whether a reverse path exists that would mirror the inverse structure
provided by integration and differentiation known from ordinary
calculus.

The availability of a corresponding derivative structure for
functionals is quite significant, as this by construction generates
spatial dependence, as indicated by $\delta/\delta\rho(x)$; see
e.g.\ Ref.~\cite{schmidt2022rmp} for details.  We can hence retrieve,
or generate, the direct correlation functional as the functional
density derivative of the intrinsic excess free energy functional:
\begin{align}
  c_1(x,[\rho]) = -\frac{\delta \beta F_\rmexc[\rho]}{\delta\rho(x)}.
  \label{EQc1definition1d}
\end{align}

While we turn to more general functional differentiation below, we
here address again the analytical case, which is useful as it reveals
the origin of the double appearance of the two spatial weighting
processes in Eqs.~\eqref{EQc1Percus}--\eqref{EQn1definition}.
Rosenfeld \cite{rosenfeld1989} introduced two weight functions
$w_0(x)$ and $w_1(x)$, which respectively describe the end points of a
particle and its interior one-dimensional ``volume'':
\begin{align}
  w_0(x) &= \frac{\delta(x-R) + \delta(x+R)}{2},
  \label{EQw0definition}\\
  w_1(x) &= \Theta(R-|x|),
  \label{EQw1definition}
\end{align}
where $\Theta(x)$ indicates the Heaviside unit step function, i.e.,
$\Theta(x\geq 0)=1$ and 0 otherwise.  The weighted densities $n_0(x)$
and $n_1(c)$, as given respectively by Eqs.~\eqref{EQn0definition} and
\eqref{EQn1definition}, can then be represented via convolution of the
respective weight function of type $\alpha=0,1$ with the density
profile according to
\begin{align}
  n_\alpha(x) &= \int dx' w_\alpha(x-x') \rho(x').
  \label{EQnalpha}
\end{align}
In more compact notation we can express \eqr{EQnalpha} as $n_\alpha(x)
= (w_\alpha \ast \rho)(x)$, where the asterisk denotes the spatial
convolution. Then the direct correlation functional is given by
\begin{align}
  c_1(x,[\rho]) 
  &= -\sum_{\alpha=0,1} (w_\alpha \ast \Phi_\alpha)(x),
  \label{EQc1PercusFormal}
\end{align}
which is an exact rewriting of the form given in \eqr{EQc1Percus}. The
coefficients $\Phi_\alpha$ are obtained as partial derivatives of the
scaled free energy density \eqref{EQPhiPercus} via $\Phi_\alpha =
\partial \Phi / \partial n_\alpha$.  This derivative structure reveals
the mechanism for the generation of the explicit forms $\Phi_0(x)$ and
$\Phi_1(x)$, as respectively given by Eqs.~\eqref{EQphi0}
and~\eqref{EQphi1}.

\subsection{Functional differentiation of direct correlations}
\label{SECdifferentiation}

While the above described use of functional differentiation in an
analytical setting might appear to be very formal and perhaps limited
in its applicability, we emphasize that the concept is indeed very
general. Given a prescribed functional of a function $\rho(x)$, the
functional derivative $\delta/\delta\rho(x)$ simply gives the gradient
of the functional with respect to a change in the input function at a
specific location $x$. 
%% The mechanism of generation of such spatial dependence is very
%% generic.

{\ms By applying the functional derivative in the present
  one-dimensional context to a given functional form of $c_1(x,
  [\rho])$, one obtains the two-body direct correlation functional and
  we recall the generic expression \eqref{EQc2AsFunctionalDerivative}
of the two-body direct correlation functional:}
\begin{align}
  c_2^\star(x,x',[\rho]) 
  &= \frac{\delta c_1^\star(x,[\rho])}{\delta\rho(x')}.
  \label{EQc2neuralFromDerivative}
\end{align}

Using the Percus version \eqref{EQc1PercusFormal} of the one-body
direct correlation functional and carrying out the functional
derivative on the right hand side of \eqr{EQc2neuralFromDerivative}
gives via an analytical calculation the following nonlocal result:
\begin{align}
  c_2(x,x',[\rho])&= -\sum_{\alpha\alpha'}
  \big(w_\alpha \ast \Phi_{\alpha\alpha'} \ast w_{\alpha'}\big)(x,x').
  \label{EQc2PercusGeneric}
\end{align}
We make the double asterisk convolution structure more explicit below.
The coefficient functions in \eqr{EQc2PercusGeneric} are obtained as
second partial derivatives via $\Phi_{\alpha\alpha'}= \partial^2
\Phi/\partial n_\alpha \partial n_{\alpha'}$. Explicitly, we have
$\Phi_{00}(x) = 0$ and the symmetry $\Phi_{01}(x) = \Phi_{10}(x)$. The
remaining terms are given by
\begin{align}
  \Phi_{01}(x) &= \frac{1}{1-n_1(x)},\\
  \Phi_{11}(x) &= \frac{n_0(x)}{[1-n_1(x)]^2}.
\end{align}

Inserting these results into \eqr{EQc2PercusGeneric} and making the
convolutions explicit yields the following expression:
\begin{align}
  c_2(x,x',[\rho]) &=
  -2 \int dx''\frac{w_0(x-x'')w_1(x'-x'')}{1-n_1(x'')}\notag\\
  & \quad - \int dx'' \frac{w_1(x-x'')n_0(x'')w_1(x'-x'')}{[1-n_1(x'')]^2}.
  \label{EQc2Percus}
\end{align}
We recall the definitions \eqref{EQw0definition} and
\eqref{EQw1definition} of the weight functions $w_0(x)$ and $w_1(x)$.
The convolution structure couples two weight functions together and
each of them has a range of $R$. Hence indeed the two-body direct
correlations are of finite range $2R=\sigma$ in the position
difference~$x-x'$~\cite{percus1976}.

{\ms While the above results for the Percus theory have been
  derived by pen-and-paper symbolic calculations, the neural
  functional is not amenable to such conventional
  techniques. Fortunately, the framework of automatic differentiation
  \cite{baydin2018autodiff} provides a powerful alternative to both
  symbolic and numerical differentiation methods, and it is a natural
  choice to consider in the context of machine learning
  \cite{chollet2017}. Via the implementation of either modified
  algebra or of computational graphs, automatic differentiation
  facilitates to obtain derivatives directly in the form of executable
  code, and crucially there is no need of any manual
  intervention. Automatic differentiation thereby is free of the
  numerical artifacts that are typical of finite difference
  schemes. The method is applicable in broad contexts, which we
  illustrate in the online tutorial
  \cite{sammueller2023neuralTutorial} by computing the Percus result
  for $c_2(x, x')$ via automatic differentiation of
  \eqr{EQc1PercusFormal} rather than by manual implementation of
  \eqr{EQc2Percus}.}

For completeness, we can recover the one-body direct correlation
functional by functional integration. We reproduce
\eqr{EQc1AsFunctionalIntegral} for the present one-dimensional
geometry:
\begin{align}
  c_1(x,[\rho]) &= 
  \int dx' \rho(x') \int_0^1 da  c_2(x,x',[\rho_a]).
  \label{EQc1FromFunctionalIntegralOneDimension}
\end{align}
On the basis of the neural representations of the direct correlation
functionals, this identity can be used to check for consistency and
for correctness of the automatic differentiation.

\subsection{Noether invariance and exchange symmetry}
\label{SECnoether}

In its standard applications Noether's theorem is used to relate
symmetries of a dynamical physical system with associated conservation
laws. Obtaining linear momentum conservation from a symmetry of the
underlying action integral is a primary example, see
e.g.\ Ref.~\cite{hermann2022topicalReview} for an introductory
presentation. Besides such deterministic applications, the Noether
theorem is currently seeing an increased use in a variety of
statistical mechanical settings \cite{baez2013markov,
  marvian2014quantum, sasa2016, sasa2019, revzen1970, budkov2022,
  brandyshev2023, bravetti2023}.

The recent statistical Noether invariance theory
\cite{hermann2021noether, hermann2022topicalReview,
  hermann2022variance, hermann2022quantum, tschopp2022forceDFT,
  sammueller2022forceDFT, sammueller2023whatIsLiquid,
  robitschko2023any, hermann2023whatIsLiquid} is based on specific
spatial displacement (``shifting'') and rotation operations. These
transformations are carried out in three-dimensional physical space
and their effect is traced back to underlying invariances on the
high-dimensional phase space and its associated thermal and
nonequilibrium ensembles.

The central statistical Noether invariance concept
\cite{hermann2021noether, hermann2022topicalReview} was demonstrated
in a range of studies, addressing the strength of force fluctuations
via their variance~\cite{hermann2022variance}, the formulation of
force-based classical density functional theory
\cite{tschopp2022forceDFT, sammueller2022forceDFT}, and the force
balance in quantum many-body systems \cite{hermann2022quantum}. The
invariance theory has led to the discovery of force-force and
force-gradient two-body correlation functions. These correlators were
shown to deliver profound insight into the microscopic spatial liquid
structure beyond the pair correlation function for a broad range of
model fluids \cite{sammueller2023whatIsLiquid,
  hermann2023whatIsLiquid}.  Noether invariance is relevant for any
thermal observable, as associated sum rules couple the given
observable to forces via very recently identified hyperforce
correlations \cite{robitschko2023any}.

The statistical Noether sum rules are exact identities that can serve
a variety of different purposes, ranging from theory building via
combination with approximate closure relations, testing for sufficient
sampling in simulation \cite{robitschko2023any}, carrying out force
sampling to improve statistical data quality and, last but not least,
testing neural functionals \cite{delasheras2023perspective,
  sammueller2023neural}. Having the latter purpose in mind, here we
describe a selection of these Noether identities.

As a fundamental property, the interparticle interaction potential
only depends on the relative particle positions and not on the
absolute particle coordinate values. Specifically, whether two
particles overlap in the one-dimensional system is unaffected by
displacing the entire microstate uniformly. This invariance against
global translation leads to associated sum rules for direct
correlation functions; we recall that the direct correlations arise
solely from the interparticle interactions and hence they are not
directly dependent on the external potential. We quote two members of
an infinite hierarchy of identities, which is originally due to
Lovett, Mou, Buff, and Wertheim \cite{lovett1976, wertheim1976}, see
Eqs.~\eqref{EQsumRulec1global} and \eqref{EQsumRulec2global} below.
We group these together with a recent curvature sum rule
\eqref{EQc2curvature} \cite{hermann2022variance}. Ultimately the
identities \eqref{EQsumRulec1global} and \eqref{EQsumRulec2global}
express the vanishing of the global interparticle force, as obtained
by summing over the interparticle forces on all particles. The three
sum rules read as follows:
\begin{align}
  \int dx \rho(x)\nabla c_1(x,[\rho]) &= 0,
  \label{EQsumRulec1global}\\
  \int dx \rho(x) \int dx' \rho(x') \nabla c_2(x,x',[\rho]) &= 0,
  \label{EQsumRulec2global}\\
  \int dx [\nabla\rho(x) ]\int dx' [\nabla'\rho(x') ] c_2(x,x',[\rho]) &=
  \notag\\ 
  -\int dx\rho(x) & \nabla\nabla c_1(x), 
  \label{EQc2curvature}
\end{align}
where in the one-dimensional system the gradient is a simple scalar
position derivative, $\nabla=d/dx$. Briefly, \eqr{EQsumRulec1global}
is obtained by noting that $F_\rmexc[\rho]=F_\rmexc[\rho_\eps]$, where
the displaced density profile is given by $\rho_\eps(\rv) =
\rho(\rv+\eps)$ with displacement vector $\eps$ (in three dimensional
systems). Building the gradient with respect to $\eps$ yields the
result $0 = \partial \beta F_\rmexc[\rho_\eps]/\partial
\eps|_{\eps=0}= \int d\rv (\delta \beta
F_\rmexc[\rho]/\delta\rho(\rv)) \nabla \rho(\rv)$, which gives
\eqr{EQsumRulec1global} upon integration by parts, resorting to the
one-dimensional geometry, and identifying the one-body direct
correlation functional via \eqr{EQc1definition1d}; for more details of
the derivation we refer the Reader to
Refs.~\cite{hermann2021noether,hermann2022topicalReview}. Equation
\eqref{EQsumRulec2global} is then obtained as the density functional
derivative of \eqr{EQsumRulec1global} and re-using
\eqr{EQsumRulec1global} to simplify the result. Equation
\eqref{EQc2curvature} is a curvature sum rule that follows from
spatial Noether invariance at second order in the global shifting
parameter $\eps$~\cite{hermann2021noether}.

Using a locally resolved shifting operation, where the displacement
$\eps(\rv)$ is local and depends on the spatial position $\rv$ and
hence constitutes a vector field (in the case of a three-dimensional
system), yields in one dimension the following position-resolved
identity:
\begin{align}
  \nabla c_1(x,[\rho]) &=
  \int dx' c_2(x,x',[\rho]) \nabla'\rho(x').
  \label{EQsumRulec1c2}
\end{align}
The left hand side has the direct interpretation of the mean
interparticle force field, expressed in units of the thermal
energy~$k_BT$. This force both acts in equilibrium and it drives the
adiabatic part of the time evolution in nonequilibrium
\cite{schmidt2022rmp}; we describe some details of the nonequilibrium
theory for time evolution in~Sec.~\ref{SECnonequilibrium}.

\begin{figure}[htb!]
  \includegraphics[page=1,width=.99\linewidth]{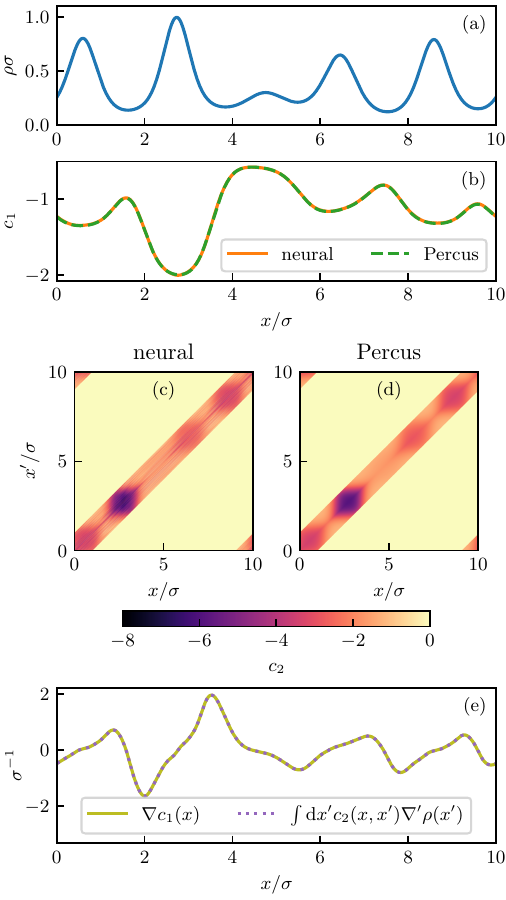}
\caption{\label{FIGfunctionalCalculus}
   Numerical results for functional calculus and Noether
   invariance. The results are shown for an exemplary oscillatory
   density profile displayed in panel (a). Results for the neural
   prediction for $c_1(x)$ are compared to numerically evaluating
   Percus' analytical direct correlation functional \eqref{EQc1Percus}
   in panel (b).  The two-body direct correlation function $c_2(x,
   x')$ as a function of $x/\sigma$ and $x'/\sigma$, as obtained from
   automatic differentiation of the neural functional is shown in
   panel (c) and compared to the result from using Percus' analytical
   expression \eqref{EQc2Percus} in panel (d).  Using the neural
   functionals, the agreement of the left and right hand side of of
   the Noether force sum rule \eqref{EQsumRulec1c2} is shown in panel
   (e).  In all cases the neural functional and Percus theories give
   numerically identical results on the scale of the respective plot.}
\end{figure}

When inserting the relationship \eqref{EQc1definition1d} of
$c_1(x,[\rho])$ to the free energy functional $F_\rmexc[\rho]$ into
the definition \eqref{EQc2neuralFromDerivative} of $c_2(x,x',[\rho])$
we obtain
\begin{align}
  c_2(x,x',[\rho]) &= 
  -\frac{\delta^2 \beta F_\rmexc[\rho]}{\delta\rho(x)\delta\rho(x')},
\end{align}
which is the one-dimensional version of the general relationship
\eqref{EQc2AsSecondFunctionalDerivative}. As the order of the two
functional derivatives is irrelevant we obtain the following exact
symmetry with respect to the exchange of the two position arguments:
\begin{align}
  c_2(x,x',[\rho]) &= c_2(x',x,[\rho]).
  \label{EQc2exchangeSymmetry}
\end{align}
When applied to the neural functional, the exchange symmetry
relationship~\eqref{EQc2exchangeSymmetry} is highly nontrivial, as the
density windows that enter the functionals on the left and on the
right hand sides differ markedly from each other, as do the
corresponding evaluation positions. That both displacement effects
cancel each other and lead to the identity
\eqref{EQc2exchangeSymmetry} is nontrivial and can serve both for
testing the quality of the neural direct correlation functional and
for demonstrating the existence of an overarching grandmother
functional $F_\rmexc[\rho]$.

In order to illustrate the theoretical structure, we display numerical
results in Fig.~\ref{FIGfunctionalCalculus}. We select a
representative oscillatory density profile, as shown in
Fig.~\ref{FIGfunctionalCalculus}(a), and take this as an input to
evaluate the one-body direct correlation functional
$c_1(x,[\rho])$. This procedure yields a specific form of the direct
correlation function $c_1(x)$, displayed in
Fig.~\ref{FIGfunctionalCalculus}(b), which belongs to the prescribed
density profile $\rho(x)$. The spatial variations of $\rho(x)$ and
$c_1(x)$ are roughly out-of-phase with each other. The nonlinear and
nonlocal nature of the functional relationship $\rho\to c_1$ is
however very apparent in the plot. The results from choosing the
neural functional or Percus' analytical one-body direct correlation
functional agree with each other to excellent accuracy. The agreement
is demonstrated in Fig.~\ref{FIGfunctionalCalculus}(b), where the two
resulting direct correlation profiles are identical on the scale of
the plot.

As laid out above above, the exchange symmetry
\eqref{EQc2exchangeSymmetry} constitutes a rigorous test for the
two-body direct correlation functional $c_2(x,x',[\rho])$. Both the
neural and the analytical functional pass with flying colours, see
Figs.~\ref{FIGfunctionalCalculus}(c) and (d) respectively, where the
symmetry of the respective ``heatmap'' graph against mirroring at the
diagonal is strikingly visible.

As a representative case for the use of a Noether sum rule as a
quantitative test for the accuracy of the neural functional methods,
we show in Fig.~\ref{FIGfunctionalCalculus}(e) the numerical results
of evaluating both sides of \eqr{EQsumRulec1c2} for the same given
density profile [shown in Fig.~\ref{FIGfunctionalCalculus}(a)]. We
find that both sides of the equation agree with high numerical
precision with each other.

\subsection{Functional integral sum rules}
\label{SECfunctionalIntegralSumRules}

{\ms We next address general identities that emerge from exploiting
  the inverse nature of functional differentiation and
  integration. For this, we recall the functional integral form
  \eqref{EQFexcFromFunctionalIntegral} of $F_\rmexc[\rho]$ and the
  functional derivative form \eqref{EQc1definition} of
  $c_1(\rho,[\rho])$, which both are central for the following
  derivations.}
That \eqr{EQFexcFromFunctionalIntegral} is the inverse of
Eq.~\eqref{EQc1definition} can be seen explicitly by functionally
differentiating \eqr{EQFexcFromFunctionalIntegral} as follows:
\begin{align}
  \frac{\delta \beta F_\rmexc[\rho]}{\delta\rho(\x)}
  =-\int d\x' \int_0^1 da \frac{\delta}{\delta\rho(\x)}
  \rho(\x')  c_1(\x',[\rho_a]).
  \label{EQc1integrateUpDifferentiateDown1}
\end{align}
We have interchanged the order of integration and functional
differentiation on the right hand side of
\eqr{EQc1integrateUpDifferentiateDown1} as these operations are
independent of each other. The functional density derivative now acts
on the product $\rho(\x')c_1(\x',[\rho_a])$ and we need to
differentiate both factors according to the product
rule. Differentiating the first factor gives the Dirac distribution,
$\delta\rho(\x')/\delta\rho(\x)=\delta(\x-\x')$. Differentiating the
second factor generates the two-body direct correlation functional
according to \eqr{EQc2neuralFromDerivative} and hence $\delta
c_1(\x',[\rho_a])/\delta\rho(\x) = a c_2(\x,\x',[\rho_a])$, where
multiplication by the scaling factor $a$ arises from the identity
$\delta/\delta\rho(\x)=a \delta/\delta(a\rho(\x))=a
\delta/\delta\rho_a(\x)$.

We can hence reformulate \eqr{EQc1integrateUpDifferentiateDown1} by
rewriting the left hand side via \eqr{EQc1definition} and expressing
the right hand side by the two separate terms. Upon multiplication by
$-1$ the result is the following functional integral identity:
\begin{align}
  c_1(\x,[\rho])=&
  \int_0^1 da c_1(\x,[\rho_a])\notag\\&
  +\int d\x' \rho(\x') \int_0^1 da a  c_2(\x,\x',[\rho_a]).
  \label{EQc1integrateUpDifferentiateDown2}
\end{align}
In the first term on the right hand side of
\eqr{EQc1integrateUpDifferentiateDown2} the position integral has
cancelled out due to the presence of the Dirac function, which leaves
over the position dependence on $\x$, as occurring in all other terms.

In order to prove \eqr{EQc1integrateUpDifferentiateDown2} and hence to
establish that indeed Eqs.~\eqref{EQFexcFromFunctionalIntegral} and
\eqref{EQc1definition} are inverse of each other, we integrate by
parts in $a$ addressing the first integral on the right hand side of
\eqr{EQc1integrateUpDifferentiateDown2}. This yields a sum of boundary
terms and an integral: $c_1(\x,[\rho])-0-\int_0^1 da a \partial
c_1(\x,[\rho_a])/\partial a$. The derivative with respect to the
parameter $a$ generates the second term in
\eqr{EQc1integrateUpDifferentiateDown2} up to the minus sign upon
carrying out the parametric derivative via $\partial/\partial a=\int
d\x'\rho(\x')\delta/\delta \rho_a(\x')$ and identifying
$c_2(\x,\x',[\rho_a])=\delta c_1(\x,[\rho_a])/\delta
\rho_a(\x')$. Hence the two integrals cancel each other. Only the
upper boundary term $c_1(\x,[\rho])$ remains, which is the left hand
side of \eqr{EQc1integrateUpDifferentiateDown2}, and hence completes
the proof.

Despite this explicit derivation via functional calculus, as both
$c_1^\star(\x,[\rho])$ and $c_2^\star(\x,\x',[\rho])$ are directly
available as neural functionals, the functional integral sum rule
\eqref{EQc1integrateUpDifferentiateDown2} provides yet again fresh
possibility for carrying our consistency and accuracy checks.

Going through the analogous chain of arguments one generation younger
leads to the following functional integral relationship between the
two- and three-body direct correlation functionals:
\begin{align}
  c_2(\x,\x',[\rho]) &= \int_0^1da c_2(\x,\x',[\rho_a])\notag\\
  & \quad +  \int d\x'' \rho(\x'')\int_0^1 da a c_3(\x,\x',\x'',[\rho_a]).
  \label{EQfunctionalSumRulec2c3}
\end{align}
The neural functional calculus allows to obtain
$c_3^\star(\x,\x',\x'',[\rho])$ via automatic generation of the
Hessian of $c_1^\star(\x,[\rho])$ \cite{sammueller2023neural}, which
elevates \eqr{EQfunctionalSumRulec2c3} beyond mere formal interest.

The structure of Eqs.~\eqref{EQc1integrateUpDifferentiateDown2} and
\eqref{EQfunctionalSumRulec2c3} expresses a general functional
relationship. When applied to the excess free energy functional itself
the result is:
\begin{align}
  \beta F_\rmexc[\rho] &= \int_0^1 da \beta F_\rmexc[\rho_a]
  \notag\\&\quad -
  \int d\x\rho(\x) \int_0^1 da a c_1(\x,[\rho_a]).
\end{align}

We furthermore demonstrate explicitly the relationship from daughter
to grandmother via functional integration of the two-body correlation
functional to obtain the excess free energy functional:
\begin{align}
  \beta F_\rmexc[\rho] &=
  -\int d\rv\rho(\rv)\int d\rv' \rho(\rv')\notag\\&\qquad\times
  \int_0^1 da \int_0^a da'
  c_2(\x,\x',[\rho_{a'}]),
  \label{EQfromDoubleFunctionalIntegral}
\end{align}
where again the scaled density profile is $\rho_{a'}(\x)=a' \rho(\x)$.
That \eqr{EQfromDoubleFunctionalIntegral} holds can be seen by
chaining together the two levels of functional integrals
\eqref{EQFexcFromFunctionalIntegral} and
\eqref{EQc1FromFunctionalIntegralOneDimension} and then simplifying
the two nested parameter integrals.  The double parametric integral in
\eqr{EQfromDoubleFunctionalIntegral} can alternatively be written with
fixed parametric boundaries as $\int_0^1da a\int_0^1 da'
c_2(\x,\x',[\rho_{aa'}])$, where the twice scaled density profile is
defined as $\rho_{aa'}(\x)=aa'\rho(\x)$.

Evans \cite{evans1992} goes further than
\eqr{EQfromDoubleFunctionalIntegral} by using the identity $\int_0^1
da \int_0^a da' f(a') = \int_0^1 da (1-a)f(a)$, which is valid for any
function $f(a)$, as can either by shown geometrically by considering
the triangle-shaped integration domain in the two-dimensional
$(a,a')$-plane or, more formally, by integration by parts.  The
identity allows to express \eqr{EQfromDoubleFunctionalIntegral} in a
form that requires to carry out only a single parametric integral:
\begin{align}
  \beta F_\rmexc[\rho] &= -\int d\rv \rho(\rv) \int d\rv' \rho(\rv')
  \notag\\&\qquad
  \times\int_0^1 da (1-a) c_2(\rv,\rv',[\rho_a]).
\end{align}
Evans \cite{evans1992} also considers more general cases where the
parameter $a$ linearly interpolates between a nontrivial initial
density profile $\rho_i(\rv)\neq 0$ and the target profile $\rho(\rv)$
via $\rho_a(\rv)=\rho_i(\rv)+a[\rho(\rv)-\rho_i(\rv)]$. In our present
description we have restricted ourselves to empty initial states,
$\rho_i(\rv)=0$, but the functional integration methodology is more
general, see Ref.~\cite{evans1992}.

Throughout we have notated the functional integrals via an outer
position integral over $\rv$ and an inner parametric integral over
$a$. This structure allows to take the common factor $\rho(\rv)$ out
of the inner integral. Standard presentations often reverse the order
of integration.  Taking the functional integral over the one-body
direct correlation functional as an example, both versions are
identical:
\begin{align}
  \int d\rv \rho(\rv) \int_0^1 da c_1(\rv,[\rho_a])
  &=
  \int_0^1 da \int d\rv \rho(\rv) c_1(\rv,[\rho_a]).
  \label{EQorderOfIntegration}
\end{align}
Our (mild) preference for the order on the left hand side of
\eqr{EQorderOfIntegration} has two reasons. i) In a numerical scheme,
where one discretizes on a grid of positions $\rv$ and of values of
$a$, the multiplication by $\rho(\rv)$ is only required to be carried
out once at each gridpoint $\rv$, when using the left hand side, not
also for every value of $a$ as on the right hand side. ii) Although
the result of the inner integral, $\int_0^1 da c_1(\rv,[\rho_a])$,
depends on the specific chosen parameterization $\rho_a(\rv)$ and is
hence not unique from the viewpoint of the entire functional, it
nevertheless constitutes a well-defined localized function of $\rv$.

\section{Nonequilibrium dynamics}
\label{SECnonequilibrium}

We have so far demonstrated that the equilibrium properties of
correlated many-body systems can be investigated on a very deep level
by using neural networks to represent the functional relationship that
are inherent in the statistical physics. The required computational
workload is thereby only quite moderate. The neural functionals that
encapsulate the nontrivial information about correlations and about
thermodynamics are lean, robust and they can be manipulated
efficiently by the neural functional calculus outlined above.

These features of the neural theory naturally lead one to wonder about
the potential applicability beyond equilibrium, i.e., to situations
where the considered system is driven by external forces such that
flow is generated. The recent nonequilibrium machine-learning method
by de las Heras {\it et al.}\ \cite{delasheras2023perspective} is
based on the dynamical one-body force balance relationship for
overdamped Brownian motion.  The required dynamical functional
dependencies are those given by power functional theory
\cite{schmidt2022rmp}. The power functional approach is formally exact
and it goes beyond dynamical density functional theory
\cite{evans1979, marconi1999, archer2004, chan2004, goddard2012prl,
  goddard2021wellposedness} in that it also captures nonequilibrium
interparticle force contributions that exceed those generated by the
free energy functional; see Refs.~\cite{schmidt2022rmp, schilling2022,
  tevrugt2020review, tevrugt2022perspective} for recent reviews. Such
genuine nonequilibrium effects include viscous and structural
nonequilibrium force fields \cite{schmidt2022rmp,
  delasheras2018velocityGradient, stuhlmueller2018prl,
  delasheras2020fourForces}, which for uniaxial compressional flow of
a three-dimensional Lennard-Jones fluid were shown to be
well-represented by a trained neural network
\cite{delasheras2023perspective}.

The neural nonequilibrium force fields were successfully compared
against analytical power functional approximations, where simple and
physically motivated semi-local dependence on both the local density
and the local velocity was shown to capture correctly the essence of
the forces that occur in the nonequilibrium situation. Together with
the exact force balance equation, this allows to predict and to design
nonequilibrium steady states \cite{delasheras2023perspective}.  The
approach offers a systematic way to go beyond dynamical density
functional theory and to address genuine nonequilibrium beyond a free
energy description. We recall studies based on dynamical density
functional theory that addressed non-equilibrium sedimentation of
colloids \cite{royall2007dynamicSedimentation}, the self-diffusion of
particles in complex fluids \cite{bier2008prl}, and the behaviour of
the van Hove two-body dynamics of colloidal Brownian hard disks
\cite{stopper2018dtpl} and of hard spheres
\cite{treffenstaedt2021dtpl, treffenstaedt2022dtpl}.

Several current statistical mechanical research threads are dedicated
to the force point of view. This includes novel force-sampling
techniques that significantly reduce the seemingly inherent
statistical noise in many-body simulation results for key quantities,
such as the density profile \cite{borgis2013,
  delasheras2018forceSampling, rotenberg2020,
  renner2023torqueSampling}. The statistical Noether invariance theory
\cite{hermann2021noether, hermann2022topicalReview,
  hermann2022variance, tschopp2022forceDFT, sammueller2022forceDFT,
  hermann2022quantum, sammueller2023whatIsLiquid, robitschko2023any,
  hermann2023whatIsLiquid} generates formal expressions for force
correlation functions very naturally.  Corresponding exact sum rules
interrelate correlations that involve forces, force gradients, and
more general observables in a hyperforce framework
\cite{robitschko2023any}.  Force-based density functional approaches
were put forward both quantum mechanically \cite{tokatly2005one,
  tokatly2005two, tokatly2007, tchenkoue2019, tarantino2021} and
classically \cite{tschopp2022forceDFT, sammueller2022forceDFT}.

We have briefly touched on the concept of forces when discussing the
direct correlation sum rule \eqref{EQsumRulec1c2}. Locally resolved
force fields are central to power functional
theory~\cite{schmidt2022rmp, schmidt2013pft, schmidt2015qpft,
  schmidt2018md} for the description of the nonequilibrium dynamics of
underlying many-body systems. The connection to the present framework
is via the locally resolved interparticle force density
$\Fv_\rmint\rt$. When expressed in correlator form, this vector field
is given as the following nonequilibrium average:
\begin{align}
  \Fv_\rmint(\rv,t) &= 
  -\avg{\sum_i \delta(\rv-\rv_i)\nabla_i u(\rv^N)}.
  \label{EQFintAsAverage}
\end{align}
The dependence on time $t$ arises as the average on the right hand
side of \eqr{EQFintAsAverage}, which is taken over the instantaneous
nonequilibrium many-body probability distribution, as given by
temporal evolution of the Smoluchowski equation for the case of
overdamped dynamics. The interparticle force density
$\Fv_\rmint(\rv,t)$ can be split into a sum of an equilibrium-like
``adiabatic'' force density $\Fv_\rmad(\rv,t)$ and a genuine
nonequilibrium ``superadiabatic'' contribution $ \Fv_\rmsup(\rv,t)$.
Making the functional dependencies explicit, as they arise in power
functional theory \cite{schmidt2022rmp}, gives the following sum of
two contributions:
\begin{align}
  \Fv_\rmint(\rv,t,[\rho,\vel]) &= 
  \Fv_\rmad(\rv,t,[\rho]) + \Fv_\rmsup(\rv,t,[\rho,\vel]).
  \label{EQFadFsupSplitting}
\end{align}
Here the functional arguments are the density profile $\rho\rt$ and
the one-body velocity field $\vel\rt = \Jv\rt/\rho\rt$, which are both
microscopically resolved in space and in time. The numerator is the
one-body current, which is given as an instantaneous nonequilibrium
average via $\Jv\rt=\langle \sum_i \delta(\rv-\rv_i)\vel_i\rangle$,
where $\vel_i(\rv^N,t)$ is the velocity of particle $i$ in the
underlying many-body overdamped Brownian dynamics.

Reference \cite{delasheras2023perspective} presents a demonstration of
the validity of the functional dependence on $\rho\rt$ and $\vel\rt$
via successful machine-learning of $\Fv_\rmint(\rv,t,[\rho,\vel])$ for
inhomogeneous nonequilibrium steady states. The strategy for
constructing the neural network is similar to that described here, but
it is based on predicting the locally resolved nonequilibrium forces
rather than the equilibrium one-body direct correlations.  One
important connection between equilibrium and nonequilibrium is given
by the adiabatic construction \cite{schmidt2022rmp} that relates
$\Fv_\rmad(\rv,t,[\rho])$ in the nonequilibrium system to an
instantaneous equilibrium system with identical density profile
$\rho(\rv,t)$. The adiabatic force field is then given as a density
functional via the standard relationship
\begin{align}
  \Fv_\rmad(\rv,t,[\rho]) = k_BT \rho(\rv,t)\nabla c_1(\rv,[\rho]),
\end{align}
where the density argument of the one-body direct correlation
functional $c_1(\rv,[\rho])$ is the instantaneous density distribution
$\rho(\rv,t)$.

For overdamped Brownian dynamics with friction constant $\gamma$, the
one-body current $\Jv\rt$ appears in the force density balance, which
is given by
\begin{align}
  \gamma \Jv\rt &= 
  -k_BT \nabla\rho\rt + \Fv_\rmint\rt + \rho\rt \fv_\rmext(\rv,t),
  \label{EQdynamicalForceDensityBalance}
\end{align}
where $ \fv_\rmext(\rv,t)$ is an external force field that acts on the
system, in general in a time- and position-dependent fashion.  The
prescription for the current is complemented by the microscopically
resolved continuity equation, $\partial\rho\rt/\partial t = -\nabla
\cdot\Jv\rt$.  Upon neglecting the superadiabatic force density in
\eqr{EQFadFsupSplitting} and hence only taking adiabatic forces into
account, i.e.\ approximating $\Fv_\rmint(\rv,t,[\rho,\vel]) \approx
\Fv_\rmad(\rv,t,[\rho])$, one arrives at the dynamical density
functional theory \cite{evans1979, marconi1999, archer2004}.  Its
inherent central approximation is hence to replace the nonequilibrium
forces by effective equilibrium forces that are obtained from the free
energy functional via the adiabatic construction
\cite{schmidt2022rmp}.

Returning to the one-dimensional geometry of the hard rod model, this
leads to the following closed approximate equation of motion for the
time-dependent density profile:
\begin{align}
  & \frac{\partial\rho(x,t)}{\partial t} =
  \notag\\
  & \quad D_0 \nabla\Big[\nabla\rho(x,t)
    -\rho(x,t)\Big( \nabla c_1(x,[\rho]) 
    + \beta f_\rmext(x,t) \Big)\Big].
  \label{EQddftOneDimensions}
\end{align}
The derivative is simply $\nabla=\partial/\partial x$ in one dimension
and the diffusion constant $D_0=k_BT/\gamma$ is the ratio of thermal
energy and the friction constant. Equation~\eqref{EQddftOneDimensions}
can be efficiently propagated in time with a simple forward Euler
algorithm and the neural representation of $c_1(x, [\rho])$ can be
used in lieu of an analytic approximation. However superadiabatic
forces, i.e.\ force contributions that go beyond the adiabatic
approximation of working with a free energy functional, are
neglected. These include viscous and structural nonequilibrium
contributions; we refer the Reader to Ref.~\cite{schmidt2022rmp} for
background and to Ref.~\cite{delasheras2023perspective} for a recent
perspective on the description of microscopic nonequilibrium dynamics
of fluids in the light of machine learning on the basis of power
functional theory.

\section{Conclusions and outlook}
\label{SECconclusions}

In conclusion we have given a detailed account of the recent neural
functional theory \cite{sammueller2023neural} for the structure and
thermodynamics of spatially inhomogeneous classical many-body
systems. The approach is based on input data obtained from Monte Carlo
simulations that provide results for averaged density
profiles. Thereby the training systems are exposed to the influence of
randomized external potentials. Based on the functional relationships
that are rigorously given by classical density functional theory, the
training data is used to construct a neural network representation of
the one-body direct correlation functional, which acts as a
fundamental ``mother'' object in the neural functional theory.

From automatic functional differentiation of the one-body direct
correlation functional follow daughter and granddaughter functionals
that represent two- and three-body direct correlation
functionals. Conversely, functional integration yields the neural
excess free energy functional, which acts as the ultimate grandmother
functional in the genealogy. We have shown that chaining together the
functional differentiation and integration operations yields exact
functional sum rules. Further exact identities are given by the
statistical mechanical Noether invariance theory
\cite{hermann2021noether, hermann2022topicalReview,
  hermann2022variance, tschopp2022forceDFT, sammueller2022forceDFT,
  hermann2022quantum, sammueller2023whatIsLiquid, robitschko2023any,
  hermann2023whatIsLiquid}, by variety of fundamental liquid state
techniques \cite{hansen2013, baus1984, evans1990, henderson1992,
  upton1998} and by functional calculus alone \cite{evans1979,
  evans1992}. We have described a selection of these sum rules in
detail and have shown how their validity can be used to carry out
consistency and accuracy checks for the different levels of mutually
related neural density functionals.

We have here in particular focused on the one-dimensional hard rod
systems for reasons of ease of data generation via simulations
\cite{sammueller2023neuralTutorial}, the availability of Percus' exact
functional \cite{percus1976}, the possibility of analytical
manipulations to be carried out, and not least the fundamental
character of this classical model \cite{tonks1936}. A
beginner-friendly interactive code tutorial is provided online
\cite{sammueller2023neuralTutorial}, together with stand-alone
documentation that describes the key strategies and the essence of the
methods that constitute the neural functional
theory~\cite{sammueller2023neural}.  We have discussed prototypical
applications for ``simulation beyond the box'', where the neural
functional is used for system sizes that outscale the dimension of the
original training box that was used to generate the underlying Monte
Carlo data~\cite{sammueller2023neural}.  We have also given an
overview of nonequilibrium methods and have emphasized the important
role of the occurring force fields and their functional dependencies.

We recall that a detailed description of the setup of the paper is
given before the start of Sec.~\ref{SECoverview}; the modular
structure of the paper invites for selective reading. An overview of
the relevant statistical mechanical concepts is given in
Sec.~\ref{SECintroduction}. The neural functional theory is described
in detail in Sec.~\ref{SECnft} and we have emphasized the important
concept of {\it local} learning, as illustrated in
Fig.~\ref{FIGnftOverview}, which facilitates very efficient network
construction and training. The neural functional approach allows to
explicitly carry out functional calculus as presented in
Sec.~\ref{SECneuralCalculus} and it is relevant for nonequilibrium as
described in Sec.~\ref{SECnonequilibrium}. We once more highlight the
availability of the online tutorial
\cite{sammueller2023neuralTutorial}, which covers all key aspects of
our study and includes a practitioner's account of automatic
differentiation and differentiable programming.

The neural functional theory is a genuine hybrid method that draws
with comparable weight from computer simulations, machine learning,
and density functional theory. The compuational and conceptual
complexities of the involved methods from each respective field are
relatively low.  Yet their combination offers a new and arguably
unprecedented view of the statistical physics of many-body systems. We
lay out in the following why the approach is interesting from the
viewpoints of each of the three constituent approaches.

From the machine-learning perspective it seems unusual to have a large
set of testable self-consistency conditions available. These
conditions stem from statistical mechanical sum rules, as they follow
e.g.\ from the Noether invariance theory, the functional
integration-differentiation structure, and exchange symmetry. Taking
the latter case as an example, that the automatic derivative of a
neural network satisfies the exchange symmetry of its two (position)
arguments is very remarkable, see the graphical demonstration of the
diagonal symmetry in Fig.~\ref{FIGfunctionalCalculus}(c).  This is a
purely structural test for the quality of the network that does not
require any independent reference data as a benchmark.  All presented
sum rules are of this type and they hence provide intrinsic
constraints, either genuinely following from the underlying
Statistical Mechanics or from mere functional calculus alone, which is
the case for the functional integration-differentiation formalism
outlined in Sec.~\ref{SECfunctionalIntegralSumRules}.  Crucially, in
our methodology the constraints are not enforced during training the
network, as is done in methods of physics-informed machine learning in
the classical density functional \cite{malpica-morales2023} and wider
\cite{karniadakis2021, jung2023piml} contexts.

From a computer simulation point of view the neural functional methods
offer a new way of designing simulation work. Instead of direct
simulation of the physical problem at hand, an intervening step of
constructing the direct correlation functional is introduced.  We have
shown that the direct correlation functional can thereby be obtained
explicitly and accurately. Rather than playing the role of a formal
object, its availability as a trained neural network facilitates
making fast and precise predictions in nontrivial situations. This
application stage of the neural functional theory requires very little
effort both in terms of the required numerical algorithmic structure
and the computational workload; we recall the illustration of the
neural functional workflow shown in Fig.~\ref{FIGworkflow}.

From a density functional perspective the neural approach is arguably
unprecedented in its degree of access to the excess free energy
functional.  We find it highly remarkable that so much of the
seemingly very abstract functional relationships and formal concepts
can be inspected and tested in computationally straightforward and
highly efficient ways. The range of these methods includes automatic
differentiation to generate direct correlation functions as well as
performant functional integration routines. The neural functional
framework offers the possibility to work numerically with exact
functional identities with great ease.  Hence the neural network
technology relieves one from the task of constructing an approximate
analytical functional and manipulating it on paper.

This leaves over the question of the status of analytical density
functionals in the light of the neural network capabilities. We have
here deliberately chosen the exact Percus functional for
one-dimensional hard rods to demonstrate how much insight can be
gleaned from the analytical manipulations; as a representative example
see the nonlinear convolutional structure of
Eqs.~\eqref{EQc1Percus}--\eqref{EQn1definition} along with the
excellent numerical comparison against the neural functional as shown
in Fig.~\ref{FIGfunctionalCalculus}(b). As the neural functional
method is not restricted to the hard core system, one can expect that
having an accurate neural functional for a given system can be of very
significant help when attempting first-principles construction of
analytical free energy functionals. After all we should make use of
the tools that van der Waals did not have at his disposal!

{\ms

In summary, in light of the progress reported in
Refs.~\cite{delasheras2023perspective, sammueller2023neural} and the
present model investigation, we anticipate that a wealth of deep
questions can be addressed from the viewpoint of the neural functional
theory, including fundamental questions of phase coexistence
\cite{binder2012} as well as the possible construction of fundamental
measure functionals \cite{leithall2011, schmidt2011internalEnergy}.
While we here have restricted ourselves to hard core systems, the
principal applicability of the neural functional theory for soft
potentials was demonstrated in Ref.~\cite{sammueller2023neural} for
(planar) inhomogeneities of the supercritical three-dimensional
Lennard-Jones fluid.  Going beyond planar geometry and addressing
spatial inhomogeneity in two or three dimensions could benefit from
the use of equivariant neural networks \cite{cohen2016, weiler2018,
  finzi2020, satorras2021, batzner2022, batzner2023, musaelian2023},
which possess the fundamental symmetry properties of Euclidean space.

For complex Hamiltonians the required amount of simulation work to
provide training data might seem as a limitation. We are however
optimistic that the subsequent efficient use of the direct correlation
functionals in the form of neural networks can by far outweigh the
training cost. Hence the application to complex models such as the
monatomic Molinero-Moore water model \cite{molinero2009, coe2022water}
might not be out of reach. Furthermore it is inspiring to think that
potential progress could be made in the treatment of dielectric
\cite{cox2020pnas} and long-ranged forces~\cite{bui2024}.

As a final note, we re-emphasize the successful application of the
neural method to nonequilibrium flow problems presented in
Ref.~\cite{delasheras2023perspective} and it is certainly very
inspiring to speculate whether this facilitates making progress
concerning questions of slow dynamics in soft matter
\cite{jung2023piml, jung2023roadmap, sammueller2023gel} and beyond
that \cite{HISML2023}.

}

\acknowledgments We thank Daniel de las Heras and Bob Evans for useful
discussions and the organizers and particants of HISML 2023
\cite{HISML2023} for inspiring feedback throughout the workshop.  This
work is supported by the German Research Foundation (DFG) via Project
No.\ 436306241.

%% \bibliographystyle{prsty}
%% \bibliography{noe}

\end{document}